\documentclass[twocolumn]{aastex631}

\usepackage{amsmath}
\usepackage{xcolor}

\usepackage[hyphens]{url} 

\usepackage{hyperref}

\begin{document}

\title{exoALMA II: Data Calibration and Imaging Pipeline}

\author[0000-0002-8932-1219]{Ryan A. Loomis}
\affiliation{National Radio Astronomy Observatory, 520 Edgemont Rd., Charlottesville, VA 22903, USA}

\author[0000-0003-4689-2684]{Stefano Facchini}
\affiliation{Dipartimento di Fisica, Universit\`a degli Studi di Milano, Via Celoria 16, 20133 Milano, Italy}

\author[0000-0002-7695-7605]{Myriam Benisty}
\affiliation{Laboratoire Lagrange, Université Côte d’Azur, CNRS, Observatoire de la Côte d’Azur, 06304 Nice, France}
\affiliation{Univ. Grenoble Alpes, CNRS, IPAG, 38000 Grenoble, France}
\affiliation{Max-Planck Institute for Astronomy (MPIA), Königstuhl 17, 69117 Heidelberg, Germany}

\author[0000-0003-2045-2154]{Pietro Curone} 
\affiliation{Dipartimento di Fisica, Universit\`a degli Studi di Milano, Via Celoria 16, 20133 Milano, Italy}
\affiliation{Departamento de Astronom\'ia, Universidad de Chile, Camino El Observatorio 1515, Las Condes, Santiago, Chile}

\author[0000-0003-1008-1142]{John~D.~Ilee} 
\affiliation{School of Physics and Astronomy, University of Leeds, Leeds, UK, LS2 9JT}

\author[0000-0002-2700-9676]{Gianni Cataldi} 
\affiliation{National Astronomical Observatory of Japan, 2-21-1 Osawa, Mitaka, Tokyo 181-8588, Japan}

\author[0000-0003-1412-893X]{Hsi-Wei Yen} 
\affiliation{Academia Sinica Institute of Astronomy \& Astrophysics, 11F of Astronomy-Mathematics Building, AS/NTU, No.1, Sec. 4, Roosevelt Rd, Taipei 10617, Taiwan}

\author[0000-0003-1534-5186]{Richard Teague}
\affiliation{Department of Earth, Atmospheric, and Planetary Sciences, Massachusetts Institute of Technology, Cambridge, MA 02139, USA}

\author[0000-0000-0000-0000]{Christophe Pinte}
\affiliation{School of Physics and Astronomy, Monash University, Clayton VIC 3800, Australia}
\affiliation{Univ. Grenoble Alpes, CNRS, IPAG, 38000 Grenoble, France}

\author[0000-0001-6947-6072]{Jane Huang} 
\affiliation{Department of Astronomy, Columbia University, 538 W. 120th Street, Pupin Hall, New York, NY, USA}

\author[0000-0002-5910-4598]{Himanshi Garg}
\affiliation{School of Physics and Astronomy, Monash University, Clayton VIC 3800, Australia}

\author[0000-0003-4039-8933]{Ryuta Orihara} 
\affiliation{College of Science, Ibaraki University, 2-1-1 Bunkyo, Mito, Ibaraki 310-8512, Japan}

\author[0000-0002-1483-8811]{Ian Czekala} 
\affiliation{School of Physics \& Astronomy, University of St. Andrews, North Haugh, St. Andrews KY16 9SS, UK}
\affiliation{Centre for Exoplanet Science, University of St. Andrews, North Haugh, St. Andrews, KY16 9SS, UK}

\author[0000-0001-9319-1296	]{Brianna Zawadzki} 
\affiliation{Department of Astronomy, Van Vleck Observatory, Wesleyan University, 96 Foss Hill Drive, Middletown, CT 06459, USA}
\affiliation{Department of Astronomy \& Astrophysics, 525 Davey Laboratory, The Pennsylvania State University, University Park, PA 16802, USA}

\author[0000-0003-2253-2270]{Sean M. Andrews}
\affiliation{Center for Astrophysics | Harvard \& Smithsonian, Cambridge, MA 02138, USA}

\author[0000-0003-1526-7587	]{David J. Wilner} 
\affiliation{Center for Astrophysics | Harvard \& Smithsonian, Cambridge, MA 02138, USA}

\author[0000-0001-7258-770X]{Jaehan Bae}
\affiliation{Department of Astronomy, University of Florida, Gainesville, FL 32611, USA}

\author[0000-0001-6378-7873]{Marcelo Barraza-Alfaro}
\affiliation{Department of Earth, Atmospheric, and Planetary Sciences, Massachusetts Institute of Technology, Cambridge, MA 02139, USA}

\author[0000-0003-4679-4072]{Daniele Fasano} 
\affiliation{Laboratoire Lagrange, Université Côte d’Azur, CNRS, Observatoire de la Côte d’Azur, 06304 Nice, France}
\affiliation{Univ. Grenoble Alpes, CNRS, IPAG, 38000 Grenoble, France}

\author[0000-0002-9298-3029]{Mario Flock} 
\affiliation{Max-Planck Institute for Astronomy (MPIA), Königstuhl 17, 69117 Heidelberg, Germany}

\author[0000-0003-1117-9213]{Misato Fukagawa} 
\affiliation{National Astronomical Observatory of Japan, 2-21-1 Osawa, Mitaka, Tokyo 181-8588, Japan}

\author[0000-0002-5503-5476]{Maria Galloway-Sprietsma}
\affiliation{Department of Astronomy, University of Florida, Gainesville, FL 32611, USA}

\author[0000-0001-8446-3026]{Andr\'es F. Izquierdo} 
\affiliation{Department of Astronomy, University of Florida, Gainesville, FL 32611, USA}
\affiliation{Leiden Observatory, Leiden University, P.O. Box 9513, NL-2300 RA Leiden, The Netherlands}
\affiliation{European Southern Observatory, Karl-Schwarzschild-Str. 2, D-85748 Garching bei M\"unchen, Germany}
\affiliation{NASA Hubble Fellowship Program Sagan Fellow}

\author[0000-0001-7235-2417]{Kazuhiro Kanagawa} 
\affiliation{College of Science, Ibaraki University, 2-1-1 Bunkyo, Mito, Ibaraki 310-8512, Japan}

\author[0000-0002-8896-9435]{Geoffroy Lesur} 
\affiliation{Univ. Grenoble Alpes, CNRS, IPAG, 38000 Grenoble, France}

\author[0000-0003-4663-0318]{Cristiano Longarini} 
\affiliation{Institute of Astronomy, University of Cambridge, Madingley Road, CB3 0HA, Cambridge, UK}
\affiliation{Dipartimento di Fisica, Universit\`a degli Studi di Milano, Via Celoria 16, 20133 Milano, Italy}

\author[0000-0002-1637-7393]{Francois Menard} 
\affiliation{Univ. Grenoble Alpes, CNRS, IPAG, 38000 Grenoble, France}

\author[0000-0002-4716-4235]{Daniel J. Price} 
\affiliation{School of Physics and Astronomy, Monash University, Clayton VIC 3800, Australia}

\author[0000-0003-4853-5736]{Giovanni Rosotti} 
\affiliation{Dipartimento di Fisica, Universit\`a degli Studi di Milano, Via Celoria 16, 20133 Milano, Italy}

\author[0000-0002-0491-143X]{Jochen Stadler} 
\affiliation{Universit\'e C\^ote d'Azur, Observatoire de la C\^ote d'Azur, CNRS, Laboratoire Lagrange, 06304 Nice, France}
\affiliation{Univ. Grenoble Alpes, CNRS, IPAG, 38000 Grenoble, France}

\author[0000-0002-3468-9577]{Gaylor Wafflard-Fernandez} 
\affiliation{Univ. Grenoble Alpes, CNRS, IPAG, 38000 Grenoble, France}

\author[0000-0002-7212-2416]{Lisa W\"olfer} 
\affiliation{Department of Earth, Atmospheric, and Planetary Sciences, Massachusetts Institute of Technology, Cambridge, MA 02139, USA}

\author[0000-0001-8002-8473	]{Tomohiro C. Yoshida} 
\affiliation{National Astronomical Observatory of Japan, 2-21-1 Osawa, Mitaka, Tokyo 181-8588, Japan}
\affiliation{Department of Astronomical Science, The Graduate University for Advanced Studies, SOKENDAI, 2-21-1 Osawa, Mitaka, Tokyo 181-8588, Japan}



\begin{abstract}
The exoALMA Large Program was designed to search for subtle kinematic deviations from Keplerian motion, indicative of embedded planets, in high angular and spectral resolution Band 7 observations of $^{12}$CO, $^{13}$CO and CS emission from protoplanetary disks. This paper summarizes the calibration and imaging pipelines used by the exoALMA collaboration. With sources ranging in diameter from $2.4\arcsec$ to $13.8\arcsec$ when probed by $^{12}$CO, multiple antennae configurations were required to maximally recover all spatial information (including the ACA for 7 sources). Combining these datasets warranted particular care in their alignment during calibration and prior to imaging, so as not to introduce spurious features that might resemble the kinematic deviations being investigated. Phase decoherence was found in several datasets, which was corrected by an iterative self-calibration procedure, and we explored the effects of the order of operations of spatial alignment, flux scaling, and self-calibration.  A number of different imaging sets were produced for the continuum and line emission, employing an iterative masking procedure that minimizes bias due to non-Keplerian motions in the disk.
\end{abstract}

\keywords{}






\section{Introduction}
\label{sec:intro}
Over the past decade, significant advances in our understanding of protoplanetary disk structures have emerged, largely thanks to cutting-edge instruments like the Atacama Large Millimeter/submillimeter Array (ALMA). Comprehensive surveys have shown that these disks are structured, with features such as rings, gaps, and spirals being nearly ubiquitous in the largest and brightest disks \citep{Andrews_2020}. This high degree of structuring is also observed in chemical studies at millimeter wavelengths \citep{Oberg_ea_2023} and scattered light imaging which traces the surface layers of disks \citep{Benisty_ea_2023}. Identifying the exact origins of such disk substructures remains a difficult task \citep{Bae_ea_2023}, as multiple mechanisms, such as planet-disk interactions, or hydrodynamical instabilities, can either act together or produce similar structures.

One promising approach to differentiate potential scenarios is to analyze the kinematic signatures of these processes within protoplanetary disks \citep{Pinte_ea_2023}. For instance, localized deviations from Keplerian motion have been detected in specific channel maps and interpreted as evidence for embedded planets \citep{Pinte_ea_2018b, Teague_ea_2018a}. On more global scales, the prominence of various (magneto-)hydrodynamical processes can be tested. For example, the kinematic signatures of vertical shear instability manifesting as axis-symmetric rings in velocity residuals \citep{Barraza-Alfaro_ea_2021}, in stark contrast to the `wiggles' of the gravitational instability \citep{Hall_ea_2020, Speedie_ea_2024}. 

In that context, the exoALMA Large Program targeted 15 bright and radially extended protoplanetary disks with the Atacama Large Millimeter/submillimeter Array (ALMA), with the goals of searching for a population of embedded planets, detecting planet-disk interactions, characterizing a broad range of dynamical features (instabilities, winds), and probing the physical conditions in these planet-forming disks \citep{Teague_exoALMA}. In order to reach these goals, the program is based on image-plane analysis of high angular (100 mas) and spectral (30.5\,kHz, 26\,m/s) resolution observations of three molecular lines: $^{12}$CO J=3-2, $^{13}$CO J=3-2 and CS J=7-6. In contrast with studies focusing on point source detection, kinematical studies require a robust measurement of the emission \textit{morphology} in each channel to detect and analyze faint small-scale variations in extended emission. This manifests as a high level requirement on the image dynamic range and image fidelity: any potential variations or artifacts in the output images introduced by our calibration or imaging procedures effectively set a noise floor on the kinematic features that can be reasonably interpreted as real.

Interferometric image fidelity is fundamentally driven by the $uv$-coverage of the observations and the quality of the data calibration. ALMA performs excellently in both regards, with very smooth $uv$-coverage for any given 12-m array configuration, the ability to request multiple array configurations and combine them, and a pipeline \citep{Hunter_ea_2023} which delivers well-calibrated data. Still, there remain both calibration and imaging challenges at the detailed level needed for the image fidelity required for a project like exoALMA. These topics have been explored extensively in the literature over the past decade, such as best practices for self-calibration \citep{Brogan_2018}, the recovery of low surface-brightness extended emission and the impact of weighting schemes \citep{Czekala_ea_2021}, spectral artifacts and high-fidelity combination of array configurations \citep{Leroy_ea_2021a}, and high dynamic range imaging and phase alignment of multiple observations \citep{Casassus_Carcamo_2022}.

\begin{figure*}
    \centering
    \includegraphics[width=\textwidth]{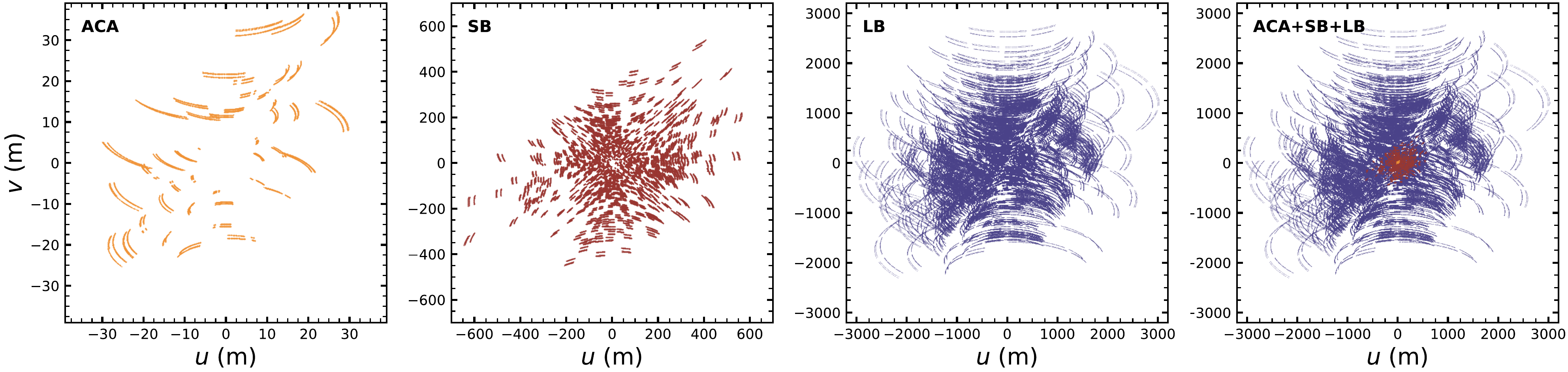}
    \caption{$uv$-coverage of the LkCa~15 dataset. The visibilities in the plot are binned in 30~s intervals. Three LB EBs were executed back-to-back, yielding optimal hour angle spanning and $uv$-coverage for such a low elevation source. Note that the final combined $uv$-coverage results in a PSF that is non-Gaussian \citep[see e.g.,][]{Czekala_ea_2021}, discussed further in \ref{sec:imaging:weighting}.}
    \label{fig:uvcoverage_LkCa15}
\end{figure*}

In this paper we present the calibration and imaging procedure adopted by the exoALMA collaboration. While many steps build upon the expertise of previous ALMA Large Programs such as DSHARP \citep{Andrews_ea_2018} and MAPS \citep{Czekala_ea_2021}, we adapted the procedure to account for the addition of Atacama Compact Array (ACA) data, the spatial alignment of multiple execution blocks containing bright, narrow line emission, and the careful assessment and correction of any phase decoherence or flux misalignment. Some of these effects, such as phase decoherence, are more pronounced at the Band 7 frequencies targeted by exoALMA, compared to the Band 6 and Band 3 observations of DSHARP and MAPS.  While many calibration and imaging techniques have been individually introduced elsewhere in the literature, we focus here on their application together in a complete pipeline and the testing done to ensure a robust set of final images, such that the kinematic features described in \citet{Izquierdo_exoALMA} and \citet{Pinte_exoALMA}, for example, can be confidently assessed as real and not due to calibration or imaging artifacts. \citet{Zawadzki_exoALMA} further assesses the confidence in our feature recovery with non-CLEAN based imaging methods, using the calibrated and aligned data described here. This paper is structured as follows. In Section~\ref{sec:observational_setup_pipeline_calibrated_data} we describe the observational setup and the ALMA pipeline calibrated data that was delivered and populates the ALMA archive. Section~\ref{sec:calibration} outlines the alignment and self-calibration processes applied to these delivered data sets. Section~\ref{sec:imaging} describes the imaging process adopted for both line and continuum images. A summary, Section~\ref{sec:summary}, concludes the paper. All scripts for performing the calibration and imaging described here can be found in the exoALMA data release.\footnote{\url{https://dataverse.harvard.edu/dataverse/exoalma}}

\section{Observational setup and pipeline-calibrated data}
\label{sec:observational_setup_pipeline_calibrated_data}

\subsection{Observational setup}

The exoALMA Large Program targeted 15 protoplanetary disks (see \cite{Teague_exoALMA} for the source list). The spectral setup comprised four spectral windows (\texttt{spw}) in Band 7. Three of them were centered on the $^{12}$CO J=3-2, $^{13}$CO J=3-2 and CS J=7-6 lines, at rest frequencies of 345.7959899\,GHz, 330.5879653\,GHz, and 342.8828503\,GHz, respectively. With a spectral sampling of 15.3\,kHz ($\sim13.5~{\rm m\,s^{-1}}$, effective velocity resolution of $26~{\rm m\,s^{-1}}$) and 3840 channels, these three \texttt{spw} each had a total bandwidth of 58.59\,MHz ($\sim51$\,km\,s$^{-1}$ at these frequencies). In order to have the targeted lines well-centered in the \texttt{spw} to cover high-velocity wings, and to allow for sufficient continuum data at both sides of the spectral lines, the sky frequencies were shifted by the line-of-sight velocities of the sources. Such velocities were derived by fitting a Keplerian model to archival line datasets (when available) using the \texttt{eddy} package \citep{eddy}, thus deriving a systemic velocity with high enough precision for our goals. The fourth \texttt{spw}, in the lower sideband, was centered at a rest frequency of 331.57\,GHz, with a maximum bandwidth of 1.875\,GHz to allow for deep continuum observations and aid in the self-calibration of the data. The spectral sampling was set to 488.5\,kHz ($440~{\rm m\,s^{-1}}$), over 3840 channels.

The observations requested a sensitivity of 3\,K in a $0.1\arcsec$ beam over one 150\,m\,s$^{-1}$ channel at the representative frequency of the $^{12}$CO J=3-2 line. With all disks spanning a Largest Angular Scale (LAS) of at least $4\arcsec$, this setup resulted in two sets of 12-m array configurations: short baselines (SB, referred to as `TM2' by ALMA) observations (close to a C-3/C-4 nominal configuration), and long baselines (LB, referred to as `TM1' by ALMA) observations (close to a C-6/C-7 nominal configuration). Low elevation targets were typically assigned more extended configurations, to compensate for the elliptical projected baselines. In these cases, phase referencing was more frequent with repetition rates ranging from 240 to 90 seconds across TM1 observations. As detailed in \citet{Teague_exoALMA}, for 7 out of the 15 sources we also requested data from the Atacama Compact Array (ACA) to recover the large scale emission of their very extended disks (i.e., those with disk diameters in $^{12}$CO above $6\arcsec$ based on archival observations).

\subsection{Cross-calibrated data}

The data published by the exoALMA collaboration were taken between October 2021 and May 2023. Dates and times of the Execution Blocks (EBs) of each Member Observing Unit Set (MOUS) are reported in Table~\ref{tab:observations} in the Appendix. Typical 12-m and 7-m EBs show an average $T_{\rm sys}\sim130 - 160\,$K, with a few exceptions. The average Precipitable Water Vapour (PWV) column was extremely good during most observations with values as low as 0.3\,mm for a few EBs. In several cases, EBs of the same MOUS were performed back-to-back, allowing for larger diversity of hour angle coverage and thus more complete $uv$-coverage (e.g., Fig.~\ref{fig:uvcoverage_LkCa15}). Even in the cases where EBs were not performed back-to-back, the snapshot 12-m $uv$-coverage of ALMA is still quite good and results in reasonable image fidelity.

The interferometric data were delivered to the exoALMA collaboration after ALMA Pipeline \citep{Hunter_ea_2023} calibration conducted by the ALMA Observatory. The calibration was performed with the software CASA \citep{CASA2022}, version 6.2.1.7. No automated self-calibration was applied to the data. The calibrated visibilities were then transferred as measurement set (MS) tables onto a server at the North American ALMA Science Center (NAASC). We performed all the subsequent calibration steps on the same server. 

Since high image fidelity of bright emission lines is imperative for kinematical studies of protoplanetary disks \citep[e.g.,][]{Disk-Dynamics_ea_2020, Pinte_ea_2023, Teague_exoALMA}, the exoALMA collaboration invested significant effort in verifying and improving the calibration of the data. These efforts build upon tests, routines and procedures developed by previous ALMA Large Programs, in particular DSHARP \citep{Andrews_ea_2018}, MAPS \citep{Oberg_ea_2021, Czekala_ea_2021} and PHANGS \citep{Leroy_ea_2021a, Leroy_ea_2021b}. 

\begin{figure*}
    \centering
    \includegraphics[width=\textwidth]{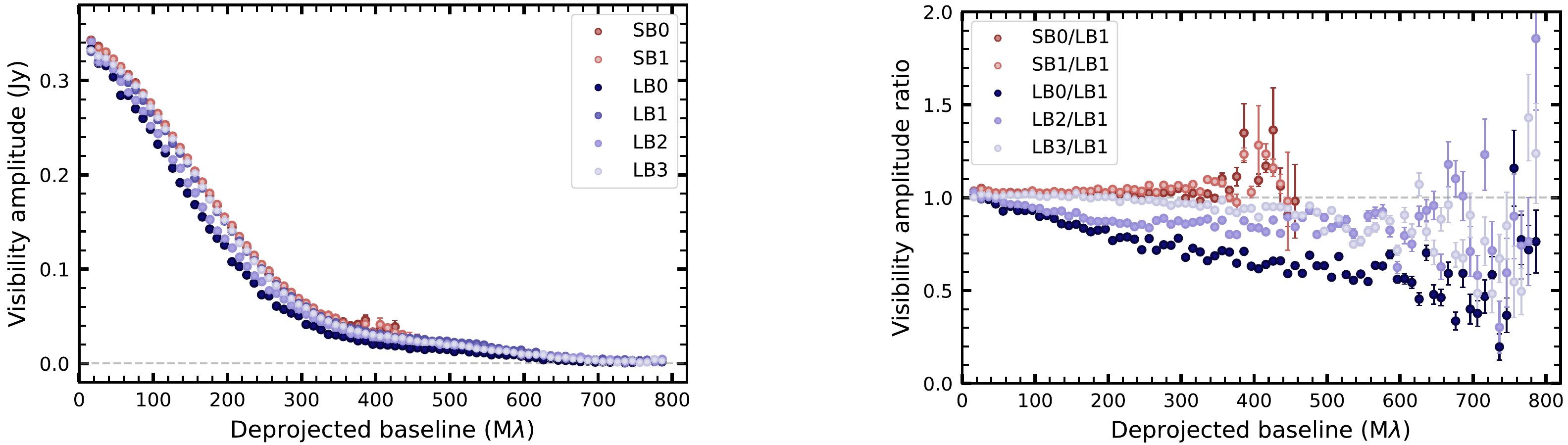}
    \caption{Visibility amplitude (left) and visibility amplitude ratio relative to LB1 (right) as a function of deprojected
    baselines for PDS~66 after self-calibration of individual EBs, alignment and flux rescaling but before group self-calibration, demonstrating misalignment between the EBs due to decorrelation from remaining phase errors.}
    \label{fig:amp_vs_bl}
\end{figure*}


Overall, the quality of the pipeline-calibrated data was very good. We manually checked the heuristics provided by the ALMA pipeline, and we did not encounter any significant problem in individual EBs. We note, however, that several EBs were assigned a semi-pass flag due to two issues. In one case (LkCa~15), the phase calibrator of the LB observations was spatially resolved, and we decided not to include these data. In the other case (J1615-3255), ALMA reported issues with maser frequency lock, causing time-variable instability in the calibration of the LO frequency, and thus the frequencies of the observations with a standard deviation of $\sim$10~kHz. By fitting a centroid to a match filtered impulse response spectrum \citep{Loomis_ea_2018} of the $^{12}$CO data on a per-scan basis, we were able to co-align the data and then match to the center frequency of unaffected execution blocks that were not set to semi-pass. The precision of this centroiding was sufficient to prevent any negative impact on the resultant images, and we did not find any structural differences between imaging J1615-3255 with or without this data (i.e. it did not introduce any artifacts).

In order to further assess the quality of the pipeline calibrated data, we also focused on three different aspects: 1) relative alignment between different EBs; 2) phase decoherence of individual EBs; 3) flux differences between different EBs. Firstly, we visually checked for continuum images of different EBs showing a relative alignment that could be improved. Secondly, we deprojected and azimuthally averaged the visibility data for disks where inclination and position angle were known from the literature. By comparing different EBs, in several cases ACA, SB and a few LB EBs suffered from significant loss of phase coherence, with the amplitudes of the averaged visibilities of different EBs having significant variations at intermediate and long baselines due to decorrelation \citep[e.g.,][see Fig.~\ref{fig:amp_vs_bl}]{Richards_ea_2022}. This was also detectable in how the visibility amplitudes varied as a function of time during the EBs, with non-Gaussian and biased scatter being present in amplitude versus time plots. Decorrelation and our strategy to improve it are further discussed in Section~\ref{sec:calibration}. Finally, it was apparent that in a few EBs, even after correcting for the loss of phase coherence, the overall flux amplitude of EBs could differ by $>4\%$. While this is still within the uncertainty of absolute flux calibration in Band 7 \citep{Francis_ea_2020, Braatz_ea_2021}, it can be adjusted to improve the final data calibration. Figures showing the issues listed above and the improvement obtained by our calibration pipeline are reported in Section~\ref{sec:calibration}.

        

\begin{figure*}
    \centering
    \includegraphics[width=2.0\columnwidth]{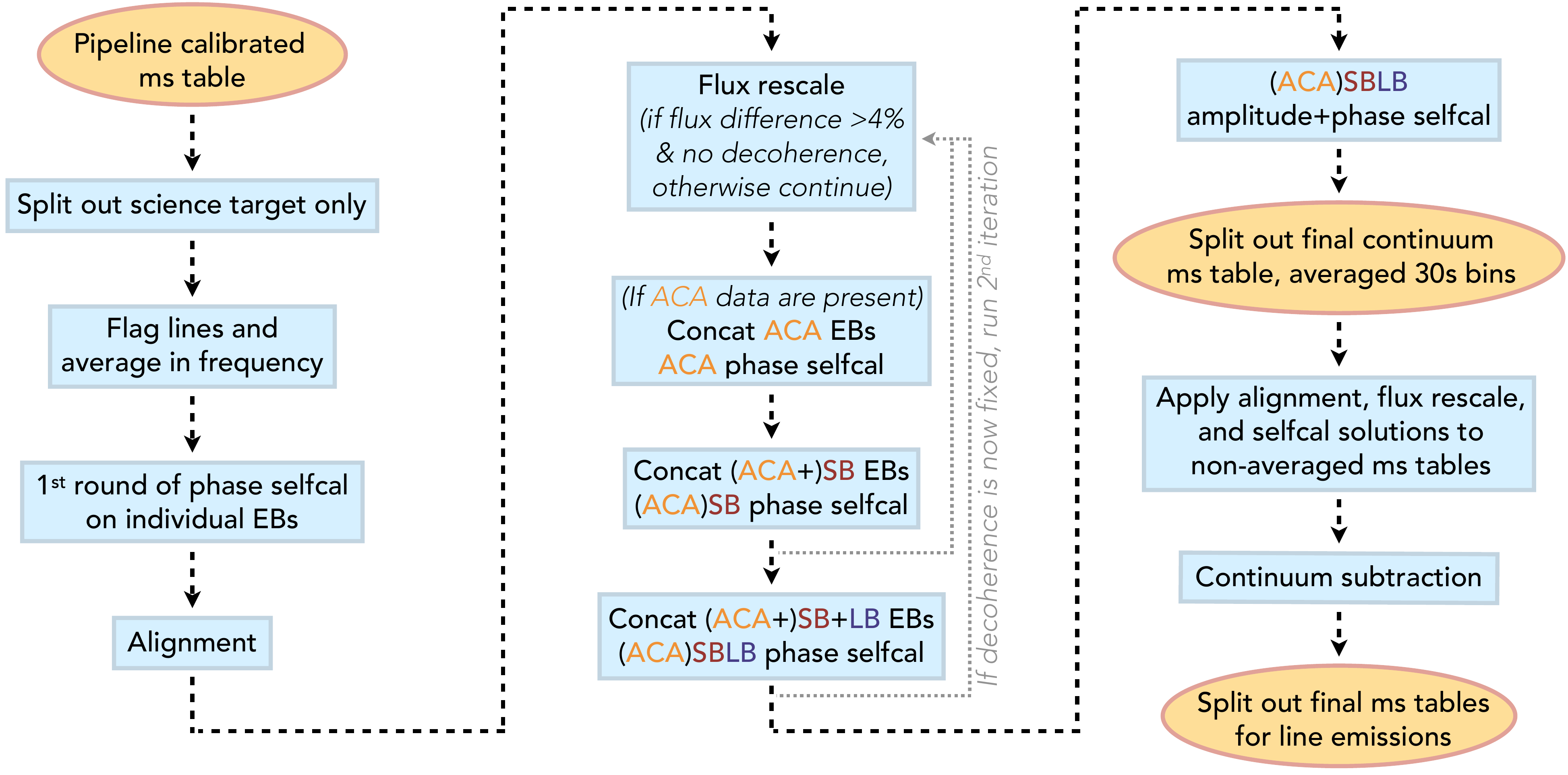}
    \caption{exoALMA calibration workflow. Individual MOUS datasets are color coded (ACA, LB, SB), and referred to with no spacing (e.g. (ACA)SBLB) when concatenated.}
    \label{fig:workflow}
\end{figure*}

\section{Self-Calibration}
\label{sec:calibration}

Self-calibration is a common procedure when processing interferometric data with a large number of antennas, given the redundancy of the number of phase and amplitude gains when compared to the number of visibilities available in any given dataset. As opposed to cross-calibration, which needs phase, amplitude and bandpass calibrators to solve the basic gain calibrator equations, a scientific source with sufficient signal-to-noise ratio is used itself as a model to compute and refine the antenna gains \citep{Brogan_2018}. While this methodology should be performed cautiously as not to introduce spurious signal into the models, particularly when applying amplitude gain solutions, it is a de facto standard technique for calibration of protoplanetary disk observations \citep[e.g.][]{Andrews_ea_2016, Andrews_ea_2018, Czekala_ea_2021}, and has recently been incorporated into the official ALMA Pipeline \citep{Hunter_ea_2023}. In this Section, we carefully describe each step we took in combining and self-calibrating the exoALMA data, motivating our choices. The workflow is summarized in Fig.~\ref{fig:workflow}. 

The exoALMA self-calibration pipeline includes custom-made routines. Many of them have been developed for the DSHARP and MAPS Large Programs \citep{Andrews_ea_2018, Oberg_ea_2021} and several were updated or modified to be used for exoALMA data. In particular, the exoALMA self-calibration pipeline includes the automated production of a large number of quality assurance figures for each step of the pipeline. All the quality assurance figures are released as part of the exoALMA data release\footnote{\url{https://dataverse.harvard.edu/dataverse/exoalma}}. Our self-calibration was performed with the same CASA version used for pipeline calibration, i.e., v6.2.1.7. While in this Section we make the effort to be comprehensive with our parameter choices in each calibration step, for the sake of readability we do not list all of them and note that these can be found in the calibration scripts released with this paper.

    

\subsection{Preparing the data for self-calibration}

We first double-checked the quality of the pipeline calibrated data by producing time-averaged amplitude versus channel plots, and time-averaged and channel-averaged amplitude versus baseline plots, for each individual \texttt{spw} of each EB. As a second step, we produced a pseudo-continuum measurement set of each EB. We did so by flagging the regions within $\pm15$\,km\,s$^{-1}$ of the systemic velocities ($v_{\rm sys}$, see Table~\ref{tab:observations}) of the sources for the $^{12}$CO, $^{13}$CO, and CS lines centers. The unflagged data were then averaged into $250$\,MHz wide channels, narrow enough to prevent frequency smearing at Band 7. With frequencies $>330$\,GHz, this choice avoids frequency smearing. A posteriori, we noticed that in a few disks, there is detectable emission of complex organic molecules (COMs)  that was not flagged (Ilee et al. in prep.). Testing the whole self-calibration pipeline with and without flagging of the COMs emission showed that there is no notable difference in the outcome. As a third pre-self-calibration heuristics, we produced a channel-averaged amplitude versus time plot of the pseudo-continuum measurement set of each EB in a narrow range of $uv$-distances where the amplitudes have high SNR. This was motivated as an additional check for the level of phase decoherence in the data, which can show up in amplitude vs time plots as `waterfalls' - narrow temporal regions of high amplitude dispersion with a one-sided bias towards zero. These waterfalls were seen in several EBs, indicating some level of decorrelation in the data. Each step of the self-calibration procedure in the next sections was conducted on the pseudo-continuum measurement sets, unless stated otherwise.

\subsection{Per-execution self-calibration}
\label{sec:selfcal_per_EB}

\begin{figure*}
    \centering
    \includegraphics[width=\textwidth]{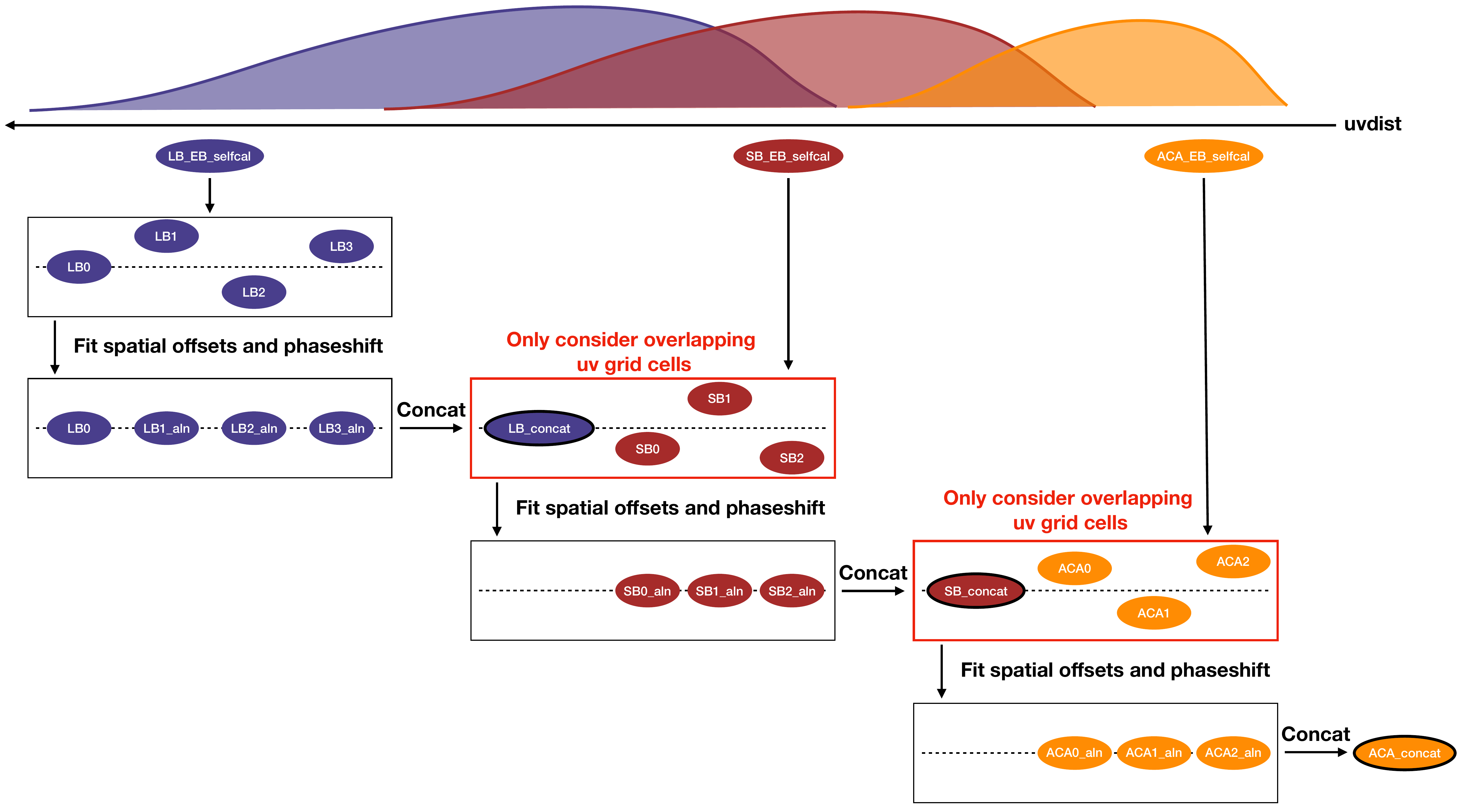}
    \caption{Schematic of the spatial alignment workflow of the exoALMA collaboration. The top histograms represent the distribution in $uv$ baseline distances for each of the three configurations considered with the longest baselines to the left. The order of operations starts in the top left and moves to the bottom right. Note that this is the inverse of our ACA-first self-calibration order of operations, with both being intentional.}
    \label{fig:alignment}
\end{figure*}

Since several EBs showed evidence of significant decorrelation, before aligning and combining different EBs within the same MOUS, we performed one round of phase-only self-calibration on each individual EB. In general for brighter protoplanetary disks such as those observed by exoALMA, there is plenty of SNR to self-calibrate single EBs by themselves. Through experimentation, this approach was shown to significantly increase the SNR of the imaged disk continuum, and therefore improve the accuracy of the relative EB alignment in the next step of our self-calibration pipeline.

We imaged the pseudo-continuum of each EB with the \texttt{tclean} task within CASA, using an elliptical mask encompassing all of the visible disk emission. For the SB and LB data, the semi-major axis of the mask was chosen manually for each disk. The semi-minor axis was reduced by a $\cos{i}$ factor, where $i$ is the inclination of the disk; the mask was then rotated by position angle (PA), also taken from literature values. For ACA EBs, we simply adopted a circular mask with a $7\arcsec$ radius. The \texttt{tclean} images were produced adopting Briggs weighting with a \texttt{robust} parameter of 0.5. Images of the SB and LB data were produced with a \texttt{multiscale} deconvolver, while ACA data used a \texttt{Hogbom} deconvolver, since our disks are not resolved in continuum by ACA standalone. The CLEAN algorithm was stopped at a 6$\sigma$ (RMS) level to avoid inserting spurious signal into the CLEAN model. The RMS was evaluated in a circular ring between $10\arcsec$ and $18\arcsec$ for ACA data, and between $4\arcsec$ and $6\arcsec$ for SB and LB data after a first CLEANing attempt. 

To self-calibrate each individual EB, model visibilities were generated for each \texttt{tclean} image to be used as model for the self-calibration. We computed antenna phase gain solutions while combining both scans and \texttt{spws} on time intervals as long as the whole EBs. As we have done for each step of the self-calibration, we applied the solutions to the data in the CASA \texttt{applycal} task using a \texttt{linearPD} interpolation to account for phase delays when we combined \texttt{spw}. A \texttt{linear} interpolation was used in the few cases where gain solutions were computed independently for each \texttt{spw}. An \texttt{applymode} = \texttt{calonly} parameter was set for the whole self-calibration pipeline, including this step, such that improved gain values were only applied where clear solutions were found, and no data was flagged due to low SNR in a given self-calibration iteration.

New CLEAN images of the data were then produced for each EB, showing improvement on the peak SNR of up to $\sim50\%$ compared to the original images, mostly due to reduced non-thermal noise. The resulting measurement sets were used as input to compute the needed relative alignment in the next step.




\subsection{Spatial alignment}

Our observations were affected by several different sources of potential spatial error, both relative (between individual executions) and absolute. First, all observations are subject to some level of pointing error as well as astrometric errors arising from phase referencing to known quasars \citep[see e.g.,][]{Fomalont_ea_2015}. For ALMA, the resulting phase center accuracy is nominally $\sim10\%$ of the clean beam major axis \citep[see ALMA Technical Handbook,][]{ALMA_THB}, although in practice it has been noticed that the scatter in real observations can be a factor of a few larger than this \citep[][and private communication, E. Fomalont (2024)]{Andrews_ea_2018}. Second, observations of protoplanetary disks as nearby objects must contend with proper motions, which may not be characterized precisely enough in the literature to ensure proper phase centering of every observation. Thus it is critical to align independent observations to a common reference point.

While the continuum emission of disks is often brightly peaked, making the image plane fit of a Gaussian to each set of observations attractive as an alignment method, the broad asymmetries and ring/gap morphologies that have been observed with ALMA \citep[e.g.,][]{Andrews_ea_2018, Francis_ea_2020} suggest that a method that is agnostic to the particular sky brightness distribution and is robust to asymmetries may be more desirable. To address this need, we developed a $uv$-plane alignment method that is similar to that proposed in \cite{Casassus_Carcamo_2022}, although developed independently and over a similar time-frame. As the two methods ended up being nearly identical in mathematical motivation, we defer the interested reader to \cite{Casassus_Carcamo_2022} for many of the method details, and instead focus here on the minor differences in our implementation of the alignment concept.

In short, for aligning any two given datasets, we first grid them to a common $uv$-grid with no additional weighting (i.e., natural weighting), and then mask out all but the grid cells which are populated in both grids, such that only data in regions of overlapping $uv$-coverage are considered (see typical $uv$-coverage in Figure \ref{fig:uvcoverage_LkCa15} and schematic at the top of Figure \ref{fig:alignment}). We then align the second dataset to the first by applying a varying positional shift (as an appropriate phase shift for each $uv$ grid cell) to the second dataset and minimizing the weighted difference between the overlapped gridded visibilities. One significant difference here between our final implementation and that in \cite{Casassus_Carcamo_2022} is that we have separated out the flux scaling alignment (see next section). The \cite{Casassus_Carcamo_2022} formulation of alignment is more generally complete, but we found potential practical motivations for separating spatial alignment and flux scaling steps. As we have removed consideration of amplitude alignment, we tested alignment implementations that only utilize phase information and minimize phase residuals rather than full gridded visibility residuals, but we found that this method had poor convergence properties. Minimizing the gridded visibility residuals resulted in similar optimal alignment shifts, but improved the convexity of the $\chi^{2}$ surface, so we retain this cost function while not introducing any flux scaling to the optimization. 

\begin{figure*}
\centering
\includegraphics[width=\textwidth]{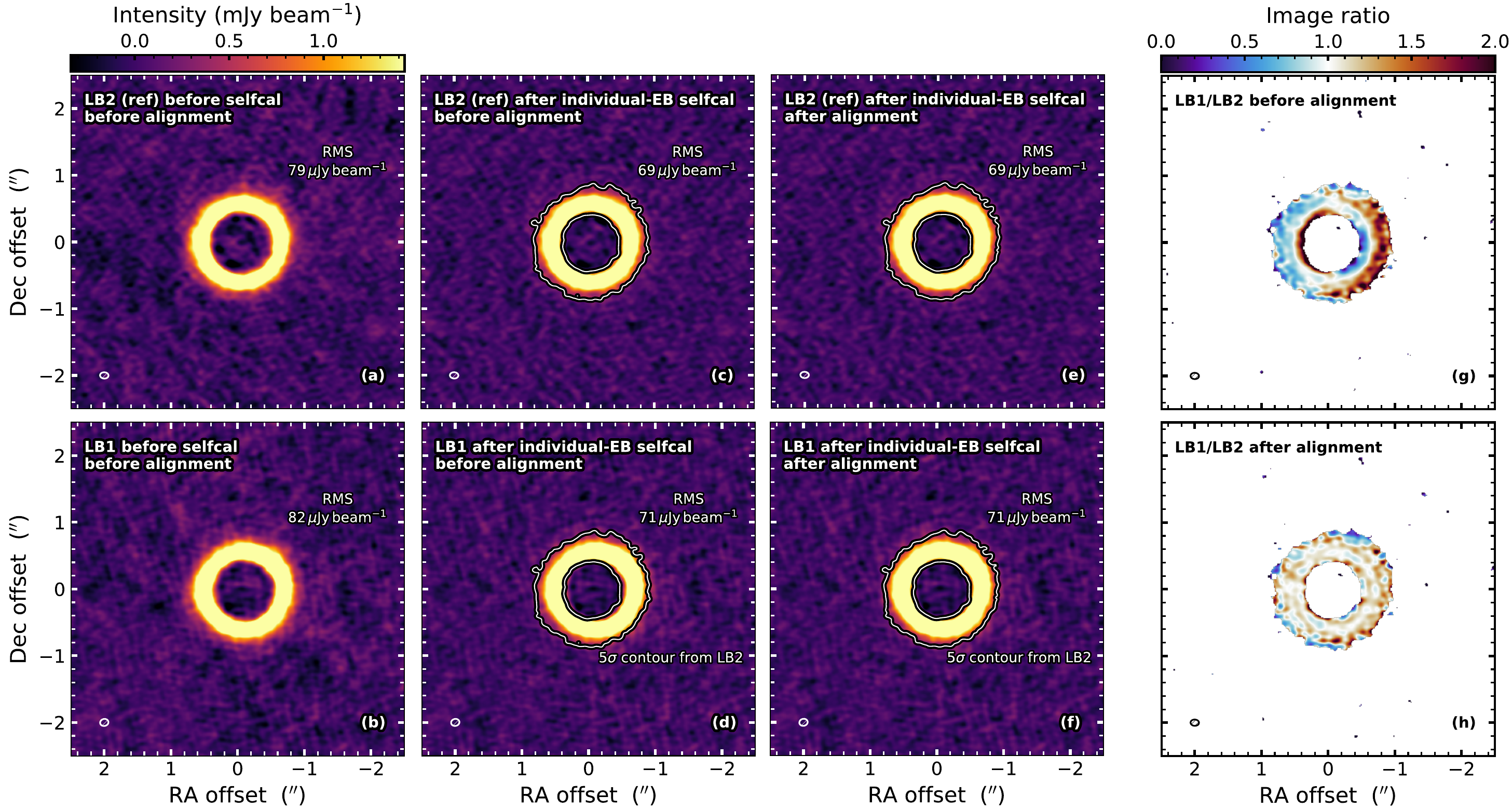}
\caption{Example of the exoALMA spatial alignment procedure on J1604 between LB1 and LB2 (the reference EB). The panels show the dust continuum images before any selfcal or alignment ($a$ and $b$), after the individual-EB selfcal ($c$ and $d$), and after the alignment ($e$ and $f$, with $e=c$ since LB2 is the reference EB and, thus, is not shifted). The contours on $c$, $d$, $e$, and $f$ indicate the $5\sigma$ level in $c$ and $e$. The RMS values were calculated in an annulus between $3\arcsec$ and $4\arcsec$ centered on the disk. The rightmost column presents the image ratio of the two EBs before and after alignment ($g=d/c$ and $h=f/e$) with a mask showing regions with $\text{SNR}>3$ in LB2. These plots highlight the successful alignment for a target that lacks a centrally peaked emission, making it unsuitable for the classical image-plane Gaussian fitting. }
\label{fig:J1604_alignment}
\end{figure*}

With this method for alignment of any two datasets in hand, we experimented with several different orders of operation for how to pick an absolute reference and align all of the datasets for a given disk. Nominally for alignment alone, this process is linear so the exact order of operations should not matter. When combining with self-calibration, however, the order of operations becomes critical as the alignment will affect the intermediate image, and therefore the model used for self-calibration. In particular with disks hosting ring and/or gap morphologies, a shift or misalignment between EBs can manifest as asymmetries which might well be interpreted as real \citep[see e.g., discussion in][]{Facchini_ea_2020}. It is important to be as careful as possible not to introduce these sorts of alignment artifacts. Figure \ref{fig:J1604_alignment} demonstrates the impact of alignment for the J1604 disk, where there is a clear artificial asymmetry visible in the disk before alignment of the two execution blocks, which is removed post-alignment.

In the end, we found that proceeding first with EB-based self-calibration (see above), then aligning the EBs, then iteratively self-calibrating combined EB's from the shortest baselines to eventually the full data yielded the most robust results. Thus the EB-based self-calibrated data forms the input for the alignment portion of our pipeline, graphically illustrated in Figure \ref{fig:alignment}. We begin by aligning the longer baseline data first, providing the most detail on the fine structure of the disk. One EB is arbitrarily taken as a reference, and the others co-aligned, and then all aligned EBs are concatenated together into a new measurement set. This concatenated LB dataset is then used as the reference for the SB data alignment (noting that only the overlapping $uv$ grid-cells between the LB and SB data are considered). The bootstrapping process then continues with the aligned and concatenated SB data used as the reference for the ACA data. The spatially co-aligned LB, SB, and ACA datasets are then ready to be flux aligned and used for the group level (e.g. multi-MOUS; SB + LB or ACA + SB) iterative self-calibration.

\subsection{Flux alignment}
\label{sec:selfcal_flux}

After the spatial alignments of the EBs, we apply a flux alignment between the EBs. Nominally this process should be possible to address simultaneously with spatial alignment, as suggested in \cite{Casassus_Carcamo_2022}, but due to the phase decoherence issues we found for some of our data, it is difficult to a priori assess 1) whether flux offsets are due to phase decoherence or absolute scaling issues, and 2) which dataset should be held as the reference. Thus we waited until after EB-based self-calibration and spatial alignment to proceed with a separate flux assessment and correction.

To determine which EBs need to have their fluxes corrected, the visibility amplitude versus deprojected baseline plots were checked for two possible flux issues. First, a constant scale factor in amplitude which can be attributed to absolute flux calibration errors. Second, a baseline-dependent offset that gets worse with longer projected baselines, indicating phase decoherence. We collectively refer to these effects as `flux offsets', as they are diagnosed via an offset between measured flux levels, but note that they are both multiplicative effects rather than additive ones. In the former case and with no evidence of decoherence, if the flux differences were larger than 4\%, a single correction was applied to bring that EB in line with the others prior to the group level self-calibration. Flux offsets below this threshold were ignored as they were empirically found not to affect the resulting image at a level that impacted our results. 

In cases where some EBs were found to exhibit phase decoherence, a modified group level self-calibration procedure was undertaken. The ACA and SB self-cal described below was performed first (to alleviate the baseline-dependent decoherence effect described above), then the flux offsets for all EBs (LB, SB, and ACA, using the SB+ACA self-calibrated data as reference) were assessed and addressed.  Finally, this flux-aligned dataset was then used for further group-level self-calibration. This process was found to be the most robust for yielding stable and consistent flux values and images.

\begin{figure*}
\centering
\includegraphics[width=\textwidth]{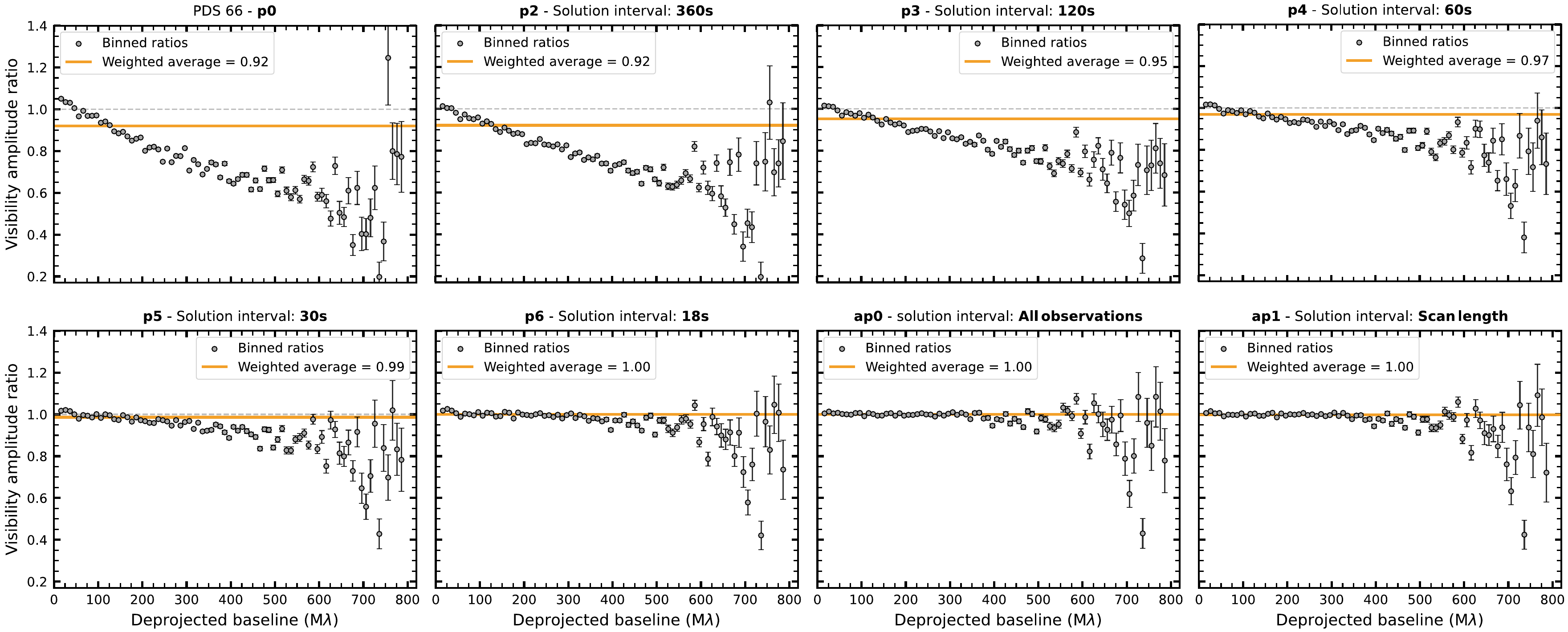}
\caption{Example of improvement of phase coherence. Visibility amplitude ratio as a function of deprojected baselines for PDS~66 between LB EB0 and LB EB1 (reference) at different rounds of SB+LB self-calibration after flux rescaling.}
\label{fig:fixed_decoherence}
\end{figure*}

\begin{figure*}
\centering
\includegraphics[width=\textwidth]{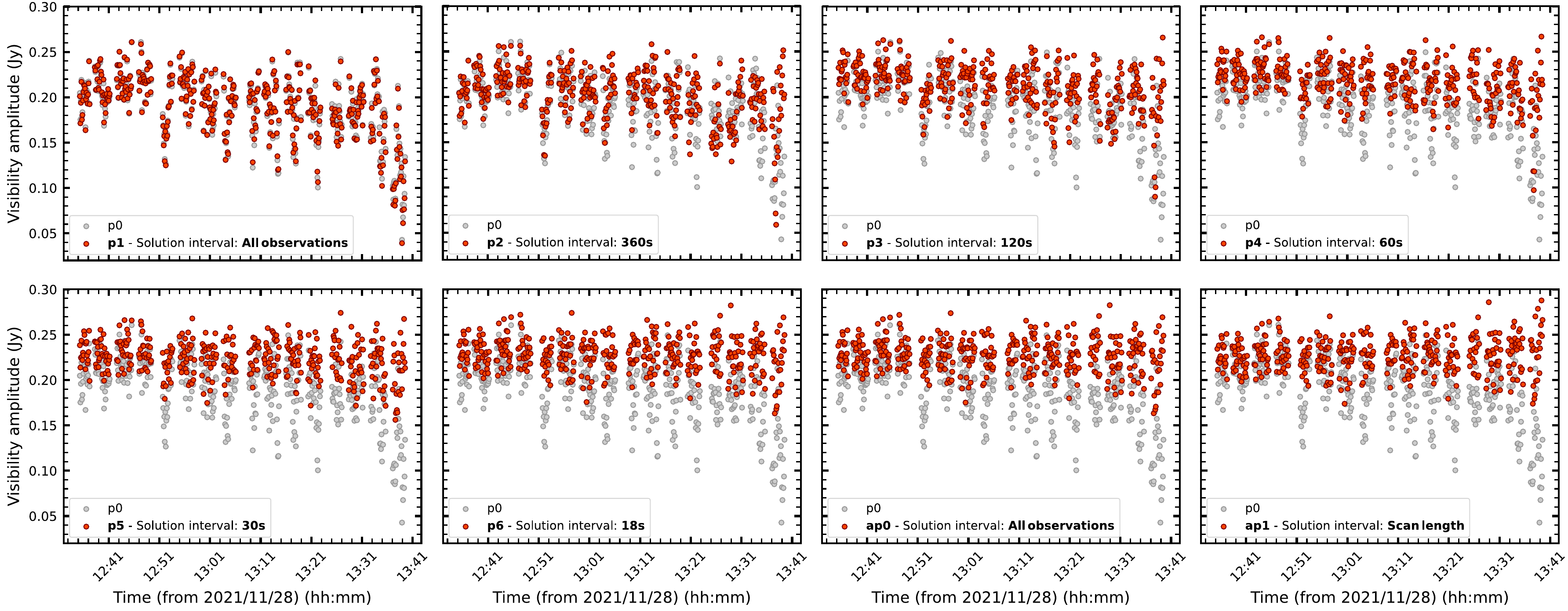}
\caption{Improvement of waterfalls in PDS 66 after flux rescaling (correlation XX, $uv$-range: 125-150 m).}
\label{fig:waterfalls}
\end{figure*}

\begin{figure*}
\centering
\includegraphics[width=1.5\columnwidth]{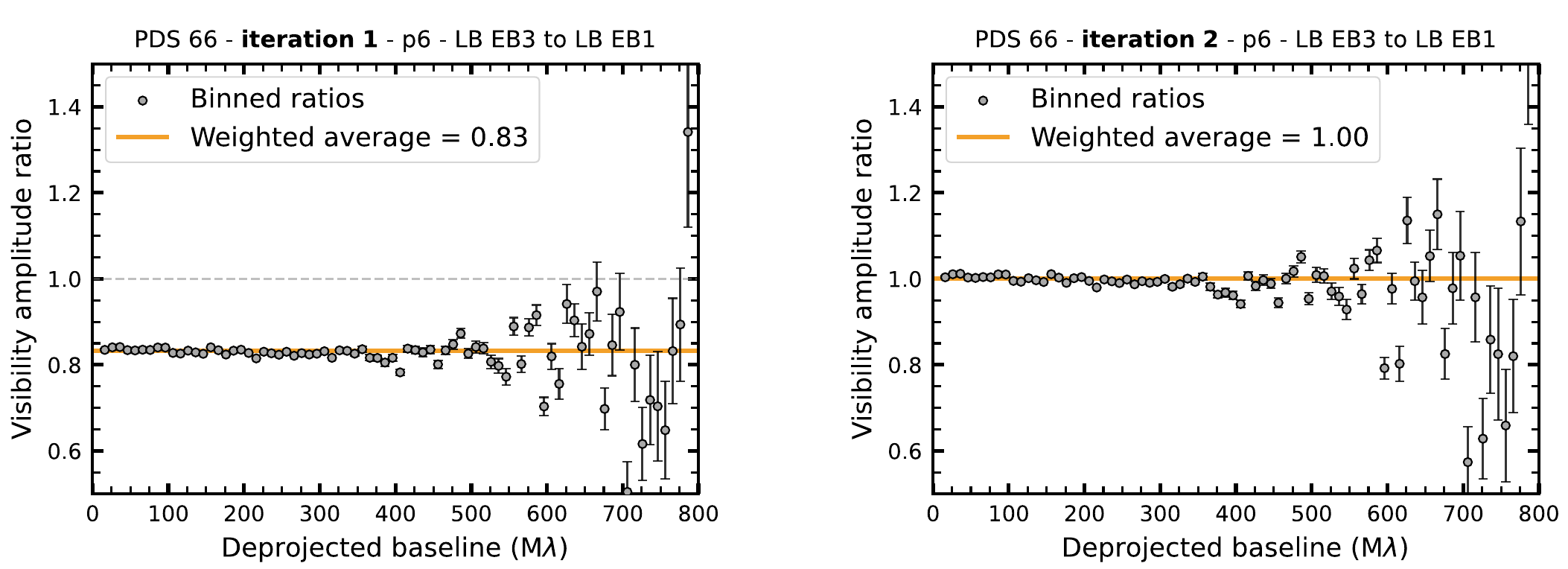}
\caption{Example of flux offset even after phase self-calibration. Visibility amplitude ratio as a function of deprojected baseline for PDS 66 at the last round of SB+LB phase selfcal of iteration 1 (left, without applying flux rescaling) and iteration 2 (right, after applying flux rescaling).}
\label{fig:flux_offset}
\end{figure*}
    
\subsection{Group level self-calibration}
\label{sec:selfcal_group-level}

Once the measurements sets were aligned both spatially and in flux, we proceeded with self-calibration of the MOUSs, as summarized in Fig.~\ref{fig:workflow}. The different subsections here represent the different steps operated in the self-calibration procedure. The strategy we used from this point forward is very similar to the one applied to the DSHARP Large Program \citep{Andrews_ea_2018}, where we progressively add longer baselines MOUSs in the self-calibration process. For all the datasets, we manually selected reference antennas that were at the center of the array and with good data quality, normally the same reference antenna as used by the ALMA pipeline and listed in the weblog.

\subsubsection{ACA phase self-calibration}
\label{sec:selfcal_ACA}

For the disks with 7-m data available, the aligned and concatenated ACA dataset was the first measurement set to be self-calibrated in phase, with two successive rounds of self-calibration. The first round was performed on EB-long intervals by combining scans, but without combining \texttt{spw} and polarizations to account for potential phase offsets between spws or polarizations. The second round of phase self-calibration was performed on 30s intervals after combining \texttt{spws}. At each step, model visibilities were created by \texttt{tclean}ing the data down to 6$\sigma$. No amplitude self-calibration was performed at this stage. For nomenclature, sub-panels in Figure \ref{fig:fixed_decoherence} refer to `p1',...,`p$n$' for the $n$-th phase self-calibration iteration, and to `ap0' and `ap1' for the two steps of amplitude and phase self-calibration iterations (see Section~\ref{sec:selfcal_LB}). 

\subsubsection{(ACA+)SB phase self-calibration - improving phase decoherence}
\label{sec:selfcal_SB}

We concatenated the aligned SB data to the self-calibrated ACA data. If ACA data are not available for the particular source, this step is the effective start of the group level self-calibration. In the calibration scripts, file labels during this step are referred to as `ACASB' (or simply `SB' when no ACA data are available). As a first step, we compared the fluxes of the ACA+SB EBs, and the ratios of the amplitudes of the deprojected visibilities (using inclination and position angle values from the literature) with respect to a reference EB, chosen for its high quality data and low phase RMS. Several EBs seemed to require an absolute flux rescaling, but we followed the workflow described in Section~\ref{sec:selfcal_flux} to decide whether to immediately rescale absolute fluxes or wait to assess the level of decorrelation of the data. We self-calibrated the phase of the ACA+SB data performing the following steps: a first round on EB-long intervals, combining \texttt{spws} but no polarizations; successive rounds progressively reducing the intervals for the gain solution (360s, 120s, 60s, 30s, 18s), combining both \texttt{spws} and polarizations. As with the ACA self-calibration, \texttt{tclean} images were computed with a cleaning threshold of 6$\sigma$. The self-calibration steps were stopped once the SNR and the noise structure of the images started degrading. Improvements on the peak SNR ratio of the data is remarkable, with improvements $>300\%$ in most cases. 

We have mentioned the loss of phase coherence as one of the major concerns on the pipeline-calibrated data. In most cases, decorrelation drastically improved during the phase self-calibration (Fig.~\ref{fig:fixed_decoherence}). Improvement was observed on time intervals ranging all the way down to $\sim18$s, highlighting how decoherence can occur on a large range of timescales, and that a frequent phase referencing can significantly improve the quality of the data, whenever possible. This is also demonstrated in Fig.~\ref{fig:waterfalls}, where the scatter and the `waterfalls' in the amplitude versus time plot in a given baseline range is significantly reduced.  

At this stage, flux offsets can be more accurately evaluated. Fig.~\ref{fig:flux_offset} shows one extreme example where, although decorrelation has been reduced by phase self-calibration, the absolute flux scales of individual EBs were still significantly different. Whenever flux offsets were $>4\%$ in at least one EB \citep[with flux offsets estimated as in][]{Andrews_ea_2018}, we went back to un-self-calibrated datasets of both ACA and SB measurement sets. We then rescaled the fluxes of the EBs showing a clear offset, and we re-performed the whole phase self-calibration process starting again from the  data after spatial alignment (see Fig.~\ref{fig:workflow}). While amplitude self-calibration could alleviate the problem, the gain solutions are computed for each antenna and so we opted instead for a uniform flux rescaling at this stage, leaving a final amplitude calibration for the end of the group self-calibration. At the end of this second iteration of the (ACA+)SB phase self-calibration, the flux offsets are verified to be all $<4\%$, as expected.

\subsubsection{(ACA+)SB+LB self-calibration - amplitude and phase self-calibration}
\label{sec:selfcal_LB}

For the final self-calibration step, we concatenated the LB EBs to the (ACA+)SB self-calibrated datasets. Our scripts refer to `ACASBLB' files (or `SBLB' files when no ACA data were taken) for this step. The approach was the same as in Sec.~\ref{sec:selfcal_SB}. We checked for decorrelation also in the LB EBs. While in most cases this was not severe, it occurred in a few datasets (e.g. in PDS~66, see Fig.~\ref{fig:fixed_decoherence}). If this was the case, we self-calibrated the data in phase, and then re-checked flux offsets. If these were $>4\%$, we went back to the beginning of the (ACA+)SB+LB self-calibration, rescaled the fluxes and re-performed the phase self-calibration. The phase self-calibration was as follows. After manually selecting reference antennas, we computed gain solutions combining \texttt{spw} for the two separate polarizations on EB-long solution intervals. We then proceeded with multiple rounds of phase self-calibration, combining polarizations, on the same solution intervals as for the (ACA+)SB data. The improvement on the peak SNR of the \texttt{tclean} images was substantial in some cases, more marginal in others, depending on the weather conditions during the observations.

\begin{figure*}
   \centering
   \includegraphics[width=\textwidth]{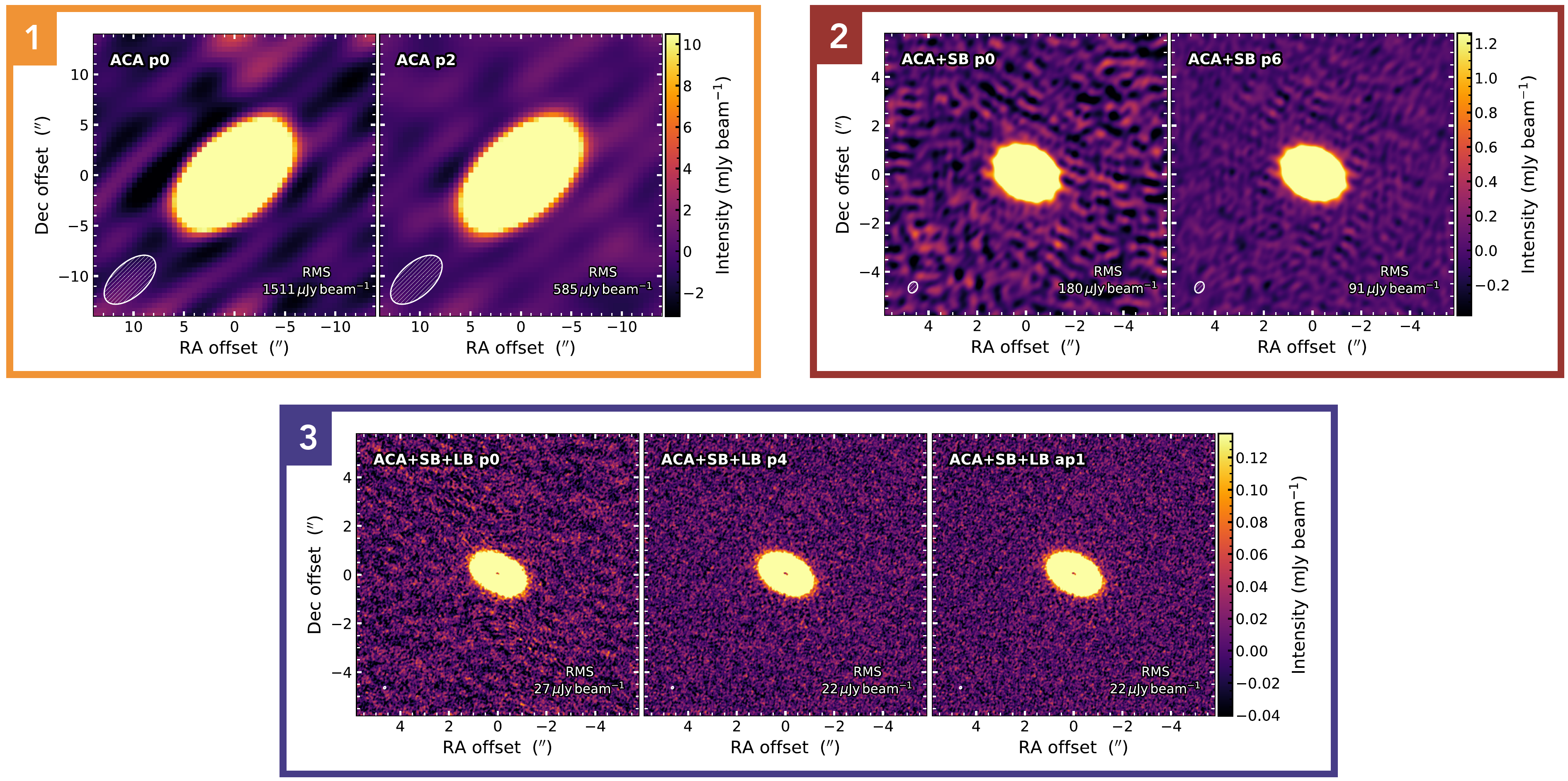}
   \caption{Demonstration of the improvements on the imaging quality of the continuum emission of LkCa~15 after the application of self-calibration on, 1) just the ACA data, 2) ACA+SB data, and 3) ACA+SB+LB data.}
   \label{fig:improvement_selfcal}
\end{figure*}

At the end of the phase self-calibration, we performed two rounds of amplitude and phase self-calibration, combining both polarizations and \texttt{spws}. A first round on EB-long intervals, and a second round on scan-long intervals. As for the phase calibration, the gain solutions were manually checked for all the EBs. While we applied a conservative ${\rm SNR}=5$ threshold on the amplitude gain solutions, we decided to flag clear solutions outliers (in particular with amplitude solutions $<0.8$ and $>1.2$). The model was produced by imaging the data with a deeper threshold than for the phase rounds, down to $1\sigma$. This is particularly important in order to include the complete flux on angular scales associated to low flux densities. The amplitude and phase self-calibration improved the peak SNR and the noise structure for several disks. An example of the overall improvement on the ACA, ACA+SB and ACA+SB+LB data during our self-calibration process is shown in Fig.~\ref{fig:improvement_selfcal}.

\subsection{Production of fiducial measurement sets}  
\label{sec:ms}

From the final self-calibration step in Section~\ref{sec:selfcal_LB}, we produced the fiducial continuum measurement set. In order to reduce its size for practical purposes, we averaged the visibilities with a 30~s time binning. We verified that time smearing did not affect our fiducial continuum images \citep{Curone_exoALMA}. We then applied the gain solutions, phase shifts, and flux alignments to the full spectral data including line emission channels. The order of operations was exactly the same as for the pseudo-continuum measurement sets, since they do not commute. The full spectral data were then binned over 30~s intervals to reduce the size of the measurement sets. Finally, continuum subtraction was applied to all the datasets, after flagging the same spectral ranges used to create the pseudo-continuum measurement sets. We used the \texttt{uvcontsub} task, with \texttt{solint}=1 and \texttt{fitorder}=1 parameters.

By the end of our self-calibration procedure we produced the following nine fiducial measurement sets for each source:
\begin{enumerate}
    \item A continuum measurement set, including all spectral windows, with a 250\,MHz spectral binning, and a 30~s time binning (e.g., \texttt{LkCa\_15\_time\_ave\_continuum.ms}).
    \item A measurement set of each of the four spectral windows with no spectral binning, and a 30~s time binning (e.g., \texttt{LkCa\_15\_ACASBLB\_no\_ave\_selfcal\_time\_ave\_12CO\\.ms}).
    \item A continuum subtracted measurement set of each of the four spectral windows with no spectral binning, and a 30~s time binning (e.g., \texttt{LkCa\_15\_ACASBLB\_no\_ave\_selfcal\_time\_ave\_12CO\\.ms.contsub}).
\end{enumerate}

These measurement sets served as input for all the images produced and analyzed by the exoALMA collaboration and are available through the exoALMA data release.



\section{Imaging}
    \label{sec:imaging}

\subsection{Weighting and $uv$-coverage effects}
    \label{sec:imaging:weighting}

Each array configuration for ALMA is designed to provide good snapshot $uv$-coverage, with a baseline distribution that yields a point spread function (PSF) that has a main lobe which is approximately Gaussian in shape and well-behaved and relatively low-lying sidelobes \citep[see ALMA Technical Handbook,][]{ALMA_THB}. Combination of these array configurations (see Figure \ref{fig:uvcoverage_LkCa15}), however, can easily result in non-Gaussian baseline distributions and therefore non-Gaussian PSFs. As first introduced in \cite{JvM_1995} and discussed extensively for protoplanetary disk observations in \cite{Czekala_ea_2021}, the effect of a non-Gaussian PSF on the final deconvolved image depends strongly on the degree of CLEANing; namely what fraction of the true sky brightness flux has been assigned to the CLEAN model. When this model is `restored' with a Gaussian CLEAN beam and combined back with the residual image, the mismatch between the flux units of Jy per CLEAN beam (Gaussian) and Jy per `dirty beam' (non-Gaussian) can result in incorrect flux measurements.

To account for these effects, an number of different image conditioning or post-processing methods have been suggested \citep{JvM_1995, Czekala_ea_2021} and debated \citep{Casassus_Carcamo_2022} in the literature. In short, these approaches fall into three categories:

\begin{enumerate}
\item Re-weight the data prior to imaging to achieve a more Gaussian baseline distribution, trading point source sensitivity for better conditioned image deconvolution and restoration.
\item Adjust the image restoration process to add the restored CLEAN model to a scaled residual image, the so-called `JvM correction' from \cite{Czekala_ea_2021}. Flux is recovered accurately and angular resolution is preserved, but point source artifacts may be introduced and point source SNR is confused by changing the residual rms. 
\item Adjust the image restoration process to use a larger restoring beam, sized to match integrated main lobe power of the original PSF. Flux is recovered accurately, point source rms is preserved, and artifacting is kept to a minimum, but angular resolution is sacrificed. 
\end{enumerate}

The MAPS Large Program imaging pipeline described in \cite{Czekala_ea_2021} took the second approach, as that program was primarily concerned with the accurate recovery of extended flux distributions at high angular resolution. \cite{Czekala_ea_2021} also outlines a CLEANing procedure that minimizes the potential artifacts introduced with this approach. There are still potential drawbacks to this method, however, particularly when the feature of interest is a point source in the presence of extended emission, such as a circumplanetary disk (CPD) \citep[e.g.][]{Benisty_ea_2021}. \citet{Casassus_Carcamo_2022} describe how the residual scaling may confuse measurements and descriptions of point source SNR in this scenario. Fundamentally, there is no catch-all perfect method for creating a restored interferometric image of sources that contain emission at a mixture of spatial scales, and any approach requires compromises.

As the exoALMA program was particularly interested in these sorts of small scale features in the presence of extended emission (both for kinematic signals in molecular line emission and potential CPDs in continuum emission), and less concerned with precise total flux recovery, we decided to avoid the `JvM correction' approach to prevent any confusion in the statistical significance of our results. Note, however, that this combined with our CLEAN masking strategy (see Section \ref{sec:line_emission}) may have a minor impact on the CS fluxes reported in \cite{Teague_exoALMA} and high velocity wings of \cite{Yoshida_exoALMA}. The third option was also found to be impractical for our purposes, as the sacrifice in angular resolution would significantly blur the small-scale kinematic features of interest. Thus we focused our efforts on the first approach, using both Briggs robust weighting and moderate $uv$-tapering to achieve PSFs that were Gaussian enough to minimize non-Gaussian effects on our kinematic analysis (explored via empirical end-to-end testing with both our imaging and analysis pipelines). We also include three different imagings of line emission with the exoALMA data release, detailed in \cite{Teague_exoALMA} and Appendix \ref{sec:appendix_images}, to allow for different views of the data. The exact weighting and imaging procedures for continuum and line emission are detailed in the following subsections.

The $\epsilon$ quantity (restoring beam power / main lobe dirty beam power) defined in \cite{Czekala_ea_2021} was measured for every image generated and recorded in the FITS image headers under the standard \texttt{HISTORY} keyword, see Section \ref{sec:line_emission}. It should be noted that these values for some images, particularly for disks with ACA observations, indicate beams that are still significantly non-Gaussian, and thus caution should be used when measuring fluxes from images with low SNR emission such as the higher resolution CS cubes. In short, the fiducial exoALMA imaging was optimized for kinematic feature analysis, and other scientific goals may require different weighting or image post-processing, which are both possible with our imaging pipeline.

\begin{figure*}
    \includegraphics[width=\textwidth]{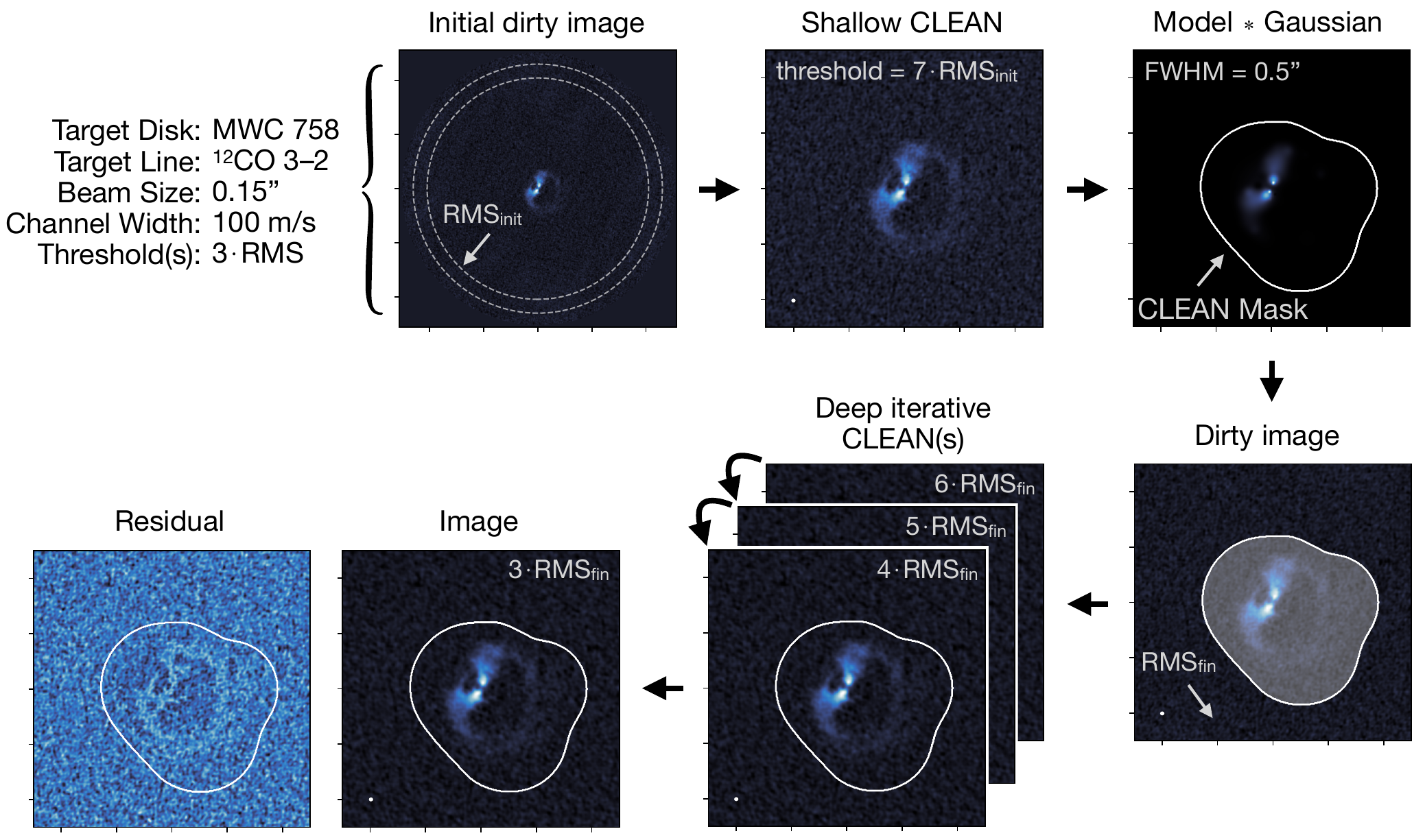}
    \caption{Representative exoALMA imaging workflow shown for a single channel of the $^{12}$CO 3--2 cube for MWC~758.}
    \label{fig:line_imaging}
\end{figure*}

\subsection{Spectral effects}
The tight coupling of spatial and kinematic structure in protoplanetary disks allows for sensitive exploration of non-Keplerian spatio-kinematic signatures, but also means that any spectral artifacts or distortions manifest as spatial distortions of signal. There are three main effects that must be taken into account.

First, any digital spectrometer will have finite sampling, and one must be careful to at minimum Nyquist sample any spectral features of interest (which correspondingly maps to minimum requirement derived from the desired spatial resolution and Keplerian parameters of the disk). This is often confused as Nyquist sampling the full `linewidth' of the disk, but should instead be thought of as (super)-Nyquist sampling the linewidth of emission from a single highly-resolved line of sight through the disk, which will be a much narrower linewidth.

Second, the spectral response function of modern interferometers is tied to both the maximum lag time of the correlator, the filter bank architecture chose, and the windowing function applied to the time domain data stream.\footnote{The spectral response function of the ALMA Baseline Correlator (BLC) is described in detail at \url{https://safe.nrao.edu/wiki/pub/Main/ALMAWindowFunctions/Note\_on_Spectral_Response.pdf}.} For ALMA, a Hann window function is applied which creates a strong covariance between channels before a variable online-binning factor is applied by the correlator, with a smaller degree of covariance remaining after this binning \citep[see e.g. Appendix C of][]{Loomis_ea_2018}. This covariance manifests as a spatio-spectral blur to the disk kinematics slightly larger than the channel spacing. For the exoALMA data, our native channel spacing ($\sim$28~m~s$^{-1}$) is significantly smaller than our smallest imaging channel width (100~m~s$^{-1}$), so the effect is quite small. 

Third, because ALMA data is taken in the topocentric frame of reference, visibilities must be regridded into a co-moving frame of reference for the source of interest (e.g. $v_{lsrk}$). The interpolation inherent to this regridding will introducing varying degrees of spectral (and therefore spatial) artifacting dependent on the method chosen. CASA defaults to linear interpolation for \texttt{tclean}.

This final effect, and minor aspects of the first two effects, are investigated in detail in the ALMA North American Development Study ``A Detailed Characterization of Spectral Regridding for the WSU Era".\footnote{A final memo for this development study is available at \url{https://science.nrao.edu/facilities/alma/science_sustainability/ALMA_Development_regridding.pdf}} For the exoALMA program, we evaluated their impact and found that our spectral sampling was high enough to mitigate their effect on accurately recovering spatio-kinematic signatures on the scale that would be introduced by exoplanetary objects. Suggestions for future improvements to observing and data reduction strategies are discussed in more detail in the development study memo.
    
\subsection{Continuum Emission}

For the continuum imaging, we followed a standard procedure using the \texttt{tclean} algorithm in CASA. We set the cell size to $0.01\arcsec$ and the image size to 1024~pixels, using the \texttt{multiscale} deconvolver with scales of [0, 8, 15, 30, 80] pixels. The \texttt{mosaic} gridder was applied to datasets with ACA data to account for the mixed primary beam patterns; otherwise, the \texttt{standard} gridder was used. No $uv$-taper or beam circularization was performed. The elliptical mask was manually defined for each disk, with a maximum semi-major axis of $3\arcsec$, and a semi-minor axis that was the semi-major axis scaled by a factor of $\cos{i}$, then rotated according to the PA of the disk. We used Briggs weighting and tested robust parameters ranging from -2 to +2 in intervals of 0.5. To determine the cleaning $\sigma$-threshold for each target and robust parameter, we ran an initial quick \texttt{tclean} iteration with $\texttt{niter}=150$ to suppress high dynamic range sidelobe effects that would otherwise cause an overestimation of the RMS, particularly at higher robust values. We then evaluated the RMS in a circular annulus between $3\arcsec$ and $4\arcsec$. A second \texttt{tclean} iteration was run with a threshold of $1\sigma$, using the RMS calculated from the first iteration. Our empirical testing found a good compromise between angular resolution and sensitivity at a robust value of -0.5, yielding continuum images with an angular resolution of ${\sim}0.09\arcsec$ and a sensitivity of ${\sim}40$~$\mu$Jy~beam$^{-1}$. We consider the continuum images with robust of -0.5 as the fiducial continuum images for the exoALMA project. The continuum images and the characterization of their substructures are presented in \cite{Curone_exoALMA}.

\subsection{Line Emission}
\label{sec:line_emission}

The line imaging pipeline used for exoALMA is generally based on the procedures used for the MAPS Large Program \citep[see][]{Czekala_ea_2021}. However, several changes were implemented to account for the complex and non-Keplerian emission morphology of our target disks and the specific science goals of the exoALMA program. Our approach to imaging of line cubes begins with the fiducial measurement sets discussed in Section \ref{sec:ms}. We show the main steps of this process performed on a single channel of the $^{12}$CO J=3-2 cube for MWC~758 in Figure \ref{fig:line_imaging} and describe them in detail below.

Input parameters for the imaging pipeline select the target measurement set (e.g. \texttt{MWC\_758\_SBLB\_no\_ave\_selfcal\_time\_ave\_12CO.ms}) which determines the disk, line, and whether or not continuum subtraction has been applied. The user also specifies the required channel spacing (e.g. 100\,m\,s$^{-1}$) and whether to image using a specific Briggs robust value or circularize the beam shape to a specified size (e.g. beam major axis = beam minor axis = 0.15\arcsec).  Beam circularization helps to remove morphological artefacts in images which may be misinterpreted as kinematic deviations that can be introduced by elliptical beams at the expense of a reduction in angular resolution along the beam minor axis. Finally, pipeline input parameters determine the CLEAN thresholds for which image products should be produced (see \cite{Teague_exoALMA} and \ref{tab:images_cube_fiducial} for description of exoALMA fiducial images). The only source specific information required is the systemic velocity of the target disk (e.g., $v_{\rm LSR} = 6.3$\,km\,s$^{-1}$, where $v_{\rm LSR}$ can be different from the $v_{\rm sys}$ used when flagging lines during calibration) and the expected width of the line emission around the systemic velocity (e.g., 10\,km\,s$^{-1}$). A pixel scale of $0\farcs025$ and an image size of $1024 \times 1024$ was chosen to provide an optimal trade off between well-sampling the synthesized beam (typically around $0\farcs15$), imaging the full primary beam (around $21\arcsec$ at 300~GHz), and limiting the resulting file size.

If a specific beam size has been requested, the beam shape is forward modeled following the approach outlined in \citet[][their Section 6.2]{Czekala_ea_2021}.  A range of Briggs robust parameters ($-2 < R < 2$) are cycled through to determine which provides a beam major axis slightly below the requested size, typically falling between 0.0 and 0.5.  An appropriate value of Gaussian $uv$-taper is then determined that will circularize the beam to the requested size.  If this is slightly below the requested size, the pipeline performs a call to \texttt{imsmooth} to reach the requested size exactly. 

With the above parameters an initial dirty image cube is created and the noise level RMS$_{\rm init}$ measured from an annulus in the outer regions (between 80--90\% of the total field of view).  A subsequent shallow CLEAN is performed down to a default threshold of $7\times\mathrm{RMS}_{\rm init}$ to capture the strongest emission.

Since one of the primary goals of the exoALMA program is to detect and characterize emission from disks that deviates from Keplerian rotation, the use of Keplerian masking during imaging is not appropriate.  Therefore, to create source specific CLEAN masks we follow a similar procedure to that adopted by the PHANGS program \citep{Leroy_ea_2021a} and the CASA auto-masking \citep{Kepley_ea_2020} whereby the CLEAN model of the previous step is convolved with a relatively wide Gaussian kernel (in practice $0\farcs5-0\farcs7$) in order to capture any spatially extended fainter  emission and then thresholded to produce a binary mask.  This process has the advantage of not relying on the disk emission being well defined in position-position-velocity space \textit{a priori}, but is instead driven by the specific emission morphology of the target source.

A final dirty image is produced and the noise level RMS$_{\rm fin}$ is measured outside of the CLEAN mask defined above.  This is followed by a final round of deep iterative CLEANing within the mask.  Pipeline input parameters can specify a range of CLEAN thresholds to produce imaging products (e.g. 6, 5, 4 and $3\times\mathrm{RMS}_{\rm fin}$).  After saving the relevant products at each requested threshold, the process is restarted from the previous CLEAN model in order to save computational time. As analyzed in detail in \citet{Zawadzki_exoALMA}, the CLEAN channel maps and non-Keplerian features in them are similarly recovered with alternative image reconstruction methods, such as Regularized Maximum Likelihood algorithms.

Finally, we note that all fiducial images produced by our pipeline contain a history of all the commands used to generate them under the standard \texttt{HISTORY} keyword.  In addition to this, we also record a range of other parameters that were used to generate the images, or measured from them, under an additional \texttt{exoALMA} keyword.  These include values for $\mathrm{RMS}_{\rm init}$, the requested CLEAN threshold, the size of the kernel used to generate the CLEAN mask, $\mathrm{RMS}_{\rm fin}$, and the peak SNR across the cube.  We also record the several properties of the restoring beam and PSF, along with the $\epsilon$ quantity (restoring beam power / main lobe dirty beam power) as defined in \cite{Czekala_ea_2021}.

\section{Summary}
\label{sec:summary}

The exoALMA Large Program collected 95 12-m Execution Blocks and 27 7-m execution blocks on a total of 15 protoplanetary disks with well-resolved gas emission. These extensive integrations have yielded some of the deepest data ever obtained at sub-mm wavelengths of planet-forming environments. This paper describes the calibration and imaging pipelines applied to the exoALMA data, carefully crafted to obtain optimal image fidelity in high spectral and angular resolution line cubes. While both pipelines have been developed leveraging the efforts of previous ALMA Large Programs \citep[e.g.][]{Andrews_ea_2018, Czekala_ea_2021, Leroy_ea_2021a}, particular attention was devoted to the following items:
\begin{itemize}
\item Several datasets suffered from loss of phase coherence, which was drastically improved by jointly self-calibrating the data. Such decorrelation changed the order of operations during the initial steps of data-combination and self-calibration as compared to other programs. Our self-calibration process was conservative, using an \texttt{applymode} = \texttt{calonly} parameter to prevent flagging of any low SNR data.
\item Due to sources exhibiting asymmetric emission morphologies without a clear central peak, all Execution Blocks for individual sources were aligned to a common phase center by minimizing the difference of gridded visibilities. Absolute flux rescaling of individual observations was applied as a separate step after accounting for decorrelation.
\item Channel maps with significantly non-Keplerian morphologies were CLEANed with the aid of masks generated by exploiting the morphology of the individual channels, rather than relying on the assumed Keplerian motion of the disk.
\end{itemize}
Our procedures, which are detailed in the main text of the paper, have provided datasets that are of sufficiently high quality to look for faint deviations from Keplerian rotation in protoplanetary disks. All the self-calibrated data, images, and scripts have been publicly released to the community to enable further improvements of calibration and analysis tools and procedures in the future.

\section*{Acknowledgments}
The authors thank the anonymous reviewer for their detailed feedback that improved the paper. This paper makes use of the following ALMA data: ADS/JAO.ALMA\#2021.1.01123.L. ALMA is a partnership of ESO (representing its member states), NSF (USA) and NINS (Japan), together with NRC (Canada), MOST and ASIAA (Taiwan), and KASI (Republic of Korea), in cooperation with the Republic of Chile. The Joint ALMA Observatory is operated by ESO, AUI/NRAO and NAOJ. The National Radio Astronomy Observatory is a facility of the National Science Foundation operated under cooperative agreement by Associated Universities, Inc. We thank the North American ALMA Science Center (NAASC) for their generous support including providing computing facilities and financial support for student attendance at workshops and publications. This work was supported by the MASSIVE HPC facility (\href{www.massive.org.au}{www.massive.org.au}). JB acknowledges support from NASA XRP grant No. 80NSSC23K1312. MB, DF, JS have received funding from the European Research Council (ERC) under the European Union’s Horizon 2020 research and innovation programme (PROTOPLANETS, grant agreement No. 101002188). Computations by JS have been performed on the `Mesocentre SIGAMM' machine, hosted by Observatoire de la Cote d’Azur. PC acknowledges support by the Italian Ministero dell'Istruzione, Universit\`a e Ricerca through the grant Progetti Premiali 2012 – iALMA (CUP C52I13000140001) and by the ANID BASAL project FB210003. SF is funded by the European Union (ERC, UNVEIL, 101076613), and acknowledges financial contribution from PRIN-MUR 2022YP5ACE. MF is supported by a Grant-in-Aid from the Japan Society for the Promotion of Science (KAKENHI: No. JP22H01274). JDI acknowledges support from an STFC Ernest Rutherford Fellowship (ST/W004119/1) and a University Academic Fellowship from the University of Leeds. Support for AFI was provided by NASA through the NASA Hubble Fellowship grant No. HST-HF2-51532.001-A awarded by the Space Telescope Science Institute, which is operated by the Association of Universities for Research in Astronomy, Inc., for NASA, under contract NAS5-26555. CL has received funding from the European Union's Horizon 2020 research and innovation program under the Marie Sklodowska-Curie grant agreement No. 823823 (DUSTBUSTERS) and by the UK Science and Technology research Council (STFC) via the consolidated grant ST/W000997/1. CP acknowledges Australian Research Council funding via FT170100040, DP18010423, DP220103767, and DP240103290. DP acknowledges Australian Research Council funding via DP18010423, DP220103767, and DP240103290. GR acknowledges funding from the Fondazione Cariplo, grant no. 2022-1217, and the European Research Council (ERC) under the European Union’s Horizon Europe Research \& Innovation Programme under grant agreement no. 101039651 (DiscEvol). FMe received funding from the European Research Council (ERC) under the European Union’s Horizon Europe research and innovation program (grant agreement No. 101053020, project Dust2Planets). H-WY acknowledges support from National Science and Technology Council (NSTC) in Taiwan through grant NSTC 113-2112-M-001-035- and from the Academia Sinica Career Development Award (AS-CDA-111-M03). GWF acknowledges support from the European Research Council (ERC) under the European Union Horizon 2020 research and innovation program (Grant agreement no. 815559 (MHDiscs)). GWF was granted access to the HPC resources of IDRIS under the allocation A0120402231 made by GENCI. Support for BZ was provided by The Brinson Foundation. TCY acknowledges support by Grant-in-Aid for JSPS Fellows JP23KJ1008. Views and opinions expressed by ERC-funded scientists are however those of the author(s) only and do not necessarily reflect those of the European Union or the European Research Council. Neither the European Union nor the granting authority can be held responsible for them. 

\appendix
\section{Details of exoALMA Observations}

Table~\ref{tab:observations} contains the relevant information of each EB used for the measurement sets and images provided with the exoALMA data release.

\startlongtable
\begin{deluxetable*}{cccccccCCC}
\tablecaption{Details of all the EBs included in the exoALMA measurement sets. \label{tab:observations}}
\tabletypesize{\footnotesize}
\tablehead{
\colhead{Source} & \colhead{Date} & \colhead{No. Ant.} & \colhead{Int} & \colhead{Baselines} & Resolution & Max. Scale & \colhead{Phase Cal.} & \colhead{Flux/Bandpass Cal.}  \\
\colhead{} & \colhead{} & \colhead{} & \colhead{\footnotesize (min)} & \colhead{\footnotesize (m)} & \colhead{\footnotesize ($\arcsec{}$)} & \colhead{\footnotesize ($\arcsec{}$)} &  &  
}
\decimals
\startdata
DM~Tau          & 2021-10-24& 8 &46  &9-35  & 3.68 &19.3  & {\rm J}0510+1800 & {\rm J}0423-0120 \\
&2021-10-24& 9 &46  &9--45  & 3.34 &19.1  & {\rm J}0510+1800 & {\rm J}0423-0120 \\
&2021-10-30& 10 &46  &9--45  & 3.34 &19.3  & 	{\rm J}0440+1437 & {\rm J}0423-0120  \\
&2022-04-14& 43 &30  &15--500  & 0.46 &5.1  & {\rm J}0510+1800 & {\rm J}0423-0120  \\
&2022-04-17& 43 &30  &15--500  & 0.45 &5.3  & {\rm J}0510+1800 & {\rm J}0423-0120  \\
&2021-11-06& 47 &46  &41--3697  & 0.06 &1.0  & 	{\rm J}0440+1437 &  {\rm J}0510+1800  \\
&2021-11-06& 47 &46  &41--3697  & 0.06 &1.0  & 	{\rm J}0440+1437 &  {\rm J}0510+1800  \\
&2021-11-06& 47 &46  &41--3697  & 0.06 &1.0  & 	{\rm J}0440+1437 &  {\rm J}0510+1800  \\
&2021-11-08& 49 &46  &41--3638  & 0.06 &1.0  & 	{\rm J}0440+1437 &  {\rm J}0510+1800  \\
AA~Tau          & 2021-11-01 & 10 & 39 & 9--45 & 3.34 & 19.3 & {\rm J}0438+3004 & {\rm J}0423-0120 \\
                & 2021-11-01 & 10 & 39 & 9--45 & 3.34 & 19.3 & {\rm J}0438+3004
                & {\rm J}0423-0120 &  \\
                & 2021-11-01 & 10 & 39 & 9--45 & 3.34 & 19.3 & {\rm J}0438+3004
                & {\rm J}0423-0120 &  \\
                & 2021-11-02 & 10 & 39 & 9--45 & 3.34 & 19.3 & {\rm J}0438+3004
                & {\rm J}0423-0120 &  \\
                & 2021-11-03 & 10 & 39 & 9--45 & 3.34 & 19.3 & {\rm J}0438+3004
                & {\rm J}0423-0120 &  \\ 
                & 2022-09-19 & 44 & 36 & 15--500 & 0.43 & 4.7 & {\rm J}0438+3004
                & {\rm J}0510+1800 &  \\ 
                & 2022-09-20 & 44 & 36 & 15--500 & 0.45 & 5.1 & {\rm J}0438+3004
                & {\rm J}0510+1800 &  \\ 
                & 2022-09-21 & 43 & 36 & 15--500 & 0.44 & 5.0 & {\rm J}0438+3004
                & {\rm J}0510+1800 &  \\ 
                & 2021-11-07 & 49 & 44 & 41--3697 & 0.06 & 0.9 & {\rm J}0435+2532
                & {\rm J}0510+1800 &  \\  
                & 2021-11-07 & 49 & 44 & 41--3697 & 0.06 & 0.9 & {\rm J}0438+3004
                & {\rm J}0510+1800 &  \\  
                & 2021-11-07 & 49 & 44 & 41--3697 & 0.06 & 0.9 & {\rm J}0435+2532
                & {\rm J}0510+1800 &  \\  
                & 2021-11-08 & 49 & 44 & 41--3638 & 0.06 & 1.0 & {\rm J}0438+3004
                & {\rm J}0510+1800 &  \\                 
LkCa~15         & 2021-10-30 & 10 & 50  & 9--45 & 3.34 & 19.3 & {\rm J}0510+1800 & {\rm J}0423-0120 \\
                & 2021-10-30 & 10 & 50  & 9--45 & 3.34 & 19.3 & {\rm J}0510+1800 & {\rm J}0423-0120   \\
                & 2021-10-31 & 9 & 50  & 9--45 & 3.29 & 13.5 & {\rm J}0510+1800 & {\rm J}0423-0120   \\
                & 2022-05-12 & 43 & 33 & 15--500 & 0.44 & 4.7 & {\rm J}0510+1800 & {\rm J}0423-0120   \\
                & 2022-09-15 & 44 & 33 & 15--782 & 0.32 & 4.2 & {\rm J}0438+3004 & {\rm J}0510+1800   \\
                & 2021-11-20 & 45 & 50 & 41--3638 & 0.06 & 1.2 & {\rm J}0435+2532 & {\rm J}0510+1800   \\ 
                & 2021-11-20 & 45 & 50 & 41--3638 & 0.06 & 1.2 & {\rm J}0435+2532 & {\rm J}0510+1800   \\
                & 2021-11-20 & 45 & 50 & 41--3638 & 0.06 & 1.2 & {\rm J}0431+1731 & {\rm J}0510+1800   \\
                & 2021-11-29 & 48 & 50 & 15--3638 & 0.08 & 1.7 & {\rm J}0435+2532 & {\rm J}0510+1800   \\
HD~34282        & 2021-10-25 & 8 & 48 & 9--35 & 3.68 & 19.3 & {\rm J}0501-0159 & {\rm J}0423-0120 \\
                & 2021-10-25 & 8 & 48 & 9--35 & 3.68 & 19.3 & {\rm J}0501-0159 & {\rm J}0423-0120   \\
                & 2021-10-25 & 8 & 48 & 9--35 & 3.68 & 19.3 & {\rm J}0542-0913 & {\rm J}0423-0120   \\
                & 2021-10-27 & 9 & 48 & 9--45 & 3.34 & 19.1 & {\rm J}0542-0913 & {\rm J}0423-0120   \\
                & 2022-01-05 & 45 & 31 & 15--784 & 0.28 & 3.6 & {\rm J}0501-0159 & {\rm J}0423-0120   \\
                & 2022-04-26 & 43 & 31 & 15--500 & 0.45 & 5.5 & {\rm J}0501-0159 & {\rm J}0423-0120   \\
                & 2021-11-09 & 47 & 48 & 41--3396 & 0.07 & 1.1 & {\rm J}0529-0519 & {\rm J}0423-0120   \\
                & 2021-11-09 & 47 & 48 & 41--3396 & 0.07 & 1.1 & {\rm J}0529-0519 & {\rm J}0423-0120   \\
                & 2021-11-10 & 48 & 48 & 45--3638 & 0.06 & 0.9 & {\rm J}0529-0519 & {\rm J}0423-0120   \\
                & 2021-11-10 & 50 & 48 & 41--3638 & 0.07 & 1.1 & {\rm J}0529-0519 & {\rm J}0423-0120   \\
                & 2021-11-14 & 52 & 48 & 41--3638 & 0.07 & 1.1 & {\rm J}0529-0519 & {\rm J}0423-0120   \\
MWC~758         & 2022-04-18 & 45 & 26 & 15--500  & 0.45 & 4.9 & {\rm J}0510+1800 & {\rm J}0423-0120 \\
                & 2022-05-14 & 44 & 26 & 15--680  & 0.37 & 4.6 & {\rm J}0510+1800 & {\rm J}0854+2006   \\
                & 2022-09-15 & 44 & 26 & 15--782  & 0.32 & 4.2 & {\rm J}0521+2112 & {\rm J}0510+1800   \\               
                & 2021-11-25 & 43 & 40 & 15--3396 & 0.07 & 1.5 & {\rm J}0521+2112 & {\rm J}0510+1800   \\
                & 2021-11-28 & 49 & 40 & 15--3638 & 0.07 & 1.6 & {\rm J}0521+2112 & {\rm J}0510+1800   \\
                & 2021-11-29 & 49 & 40 & 15--3638 & 0.08 & 1.7 & {\rm J}0521+2112 & {\rm J}0510+1800  \\
                & 2021-12-01 & 49 & 40 & 15--3638 & 0.08 & 1.8 & {\rm J}0521+2112 & {\rm J}0510+1800   \\
                & 2021-12-02 & 46 & 40 & 15--2617 & 0.09 & 1.7 & {\rm J}0521+2112 & {\rm J}0510+1800   \\
CQ~Tau          & 2022-04-17 & 43 & 36 & 15--500 & 0.45 & 5.3 & {\rm J}0510+1800 & {\rm J}0423-0120  \\
                & 2022-04-19 & 41 & 36 & 15--500 & 0.45 & 5.3 & {\rm J}0510+1800 & {\rm J}0423-0120   \\                
                & 2021-11-08 & 46 & 44 & 41--3638 & 0.06 & 1.0 & {\rm J}0550+2326 & {\rm J}0510+1800   \\
                & 2021-11-11 & 49 & 44 & 41--3638 & 0.07 & 1.1 & {\rm J}0550+2326 & {\rm J}0510+1800   \\
                & 2021-11-12 & 46 & 44 & 41--3638 & 0.06 & 1.0 & {\rm J}0550+2326 & {\rm J}0510+1800   \\
                & 2021-11-14 & 52 & 44 & 41--3638 & 0.07 & 1.1 & {\rm J}0550+2326 & {\rm J}0510+1800   \\
SY~Cha          & 2022-01-29 & 44 & 42  & 15--500  & 0.44 & 5.1 & {\rm J}1058-8003 & {\rm J}0538-4405  \\ 
                & 2022-01-30 & 44 & 42  & 15--500  & 0.44 & 5.0 & {\rm J}1058-8003 & {\rm J}0538-4405   \\ 
                & 2021-11-25 & 43 & 47  & 15--3396 & 0.07 & 1.5 & {\rm J}1058-8003 & {\rm J}1427-4206   \\
                & 2021-11-25 & 43 & 47  & 15--3396 & 0.07 & 1.5 & {\rm J}1058-8003 & {\rm J}0519-4546   \\ 
                & 2021-11-28 & 49 & 47  & 15--3638 & 0.07 & 1.6 & {\rm J}1058-8003 & {\rm J}0519-4546    \\
                & 2021-11-29 & 43 & 47  & 15--3638 & 0.08 & 1.7 & {\rm J}1058-8003 & {\rm J}0519-4546  \\      
PDS~66          & 2022-04-16 & 48 & 27 & 15--500 & 0.45 & 4.9 & {\rm J}1147-6753 & {\rm J}1427-4206  \\
                & 2022-04-16 & 48 & 27 & 15--500 & 0.45 & 4.9 & {\rm J}1147-6753 & {\rm J}1427-4206   \\
                & 2021-11-28 & 47 & 42 & 15--3638 & 0.08 & 1.6 & {\rm J}1424-6807 & {\rm J}1427-4206   \\
                & 2021-12-02 & 47 & 42 & 15--2617 & 0.09 & 1.7 & {\rm J}1308-6707 & {\rm J}1427-4206   \\
                & 2021-12-02 & 47 & 42 & 15--2617 & 0.09 & 1.7 & {\rm J}1308-6707 & {\rm J}1427-4206   \\
                & 2022-07-12 & 44 & 42 & 15--2617 & 0.1 & 1.7 & {\rm J}1424-6807 & {\rm J}1427-4206   \\
HD~135344B      & 2022-01-04 & 47 & 32  & 15--977  & 0.27 & 3.5 & {\rm J}1457-3539 & {\rm J}1427-4206  \\
                & 2022-04-19 & 48 & 32  & 15--500  & 0.45 & 4.9 & {\rm J}1454-3747 & {\rm J}1427-4206   \\
                & 2022-07-24 & 43 & 48  & 15--2617  & 0.09 & 1.6 & {\rm J}1454-3747 & {\rm J}1427-4206   \\
                & 2022-07-25 & 41 & 48  & 15--2617  & 0.09 & 1.6 & {\rm J}1454-3747 & {\rm J}1427-4206   \\
                & 2022-07-25 & 41 & 48  & 15--2617  & 0.09 & 1.6 & {\rm J}1454-3747 & {\rm J}1427-4206   \\
                & 2022-07-25 & 43 & 48  & 15--2617  & 0.09 & 1.7 & {\rm J}1454-3747 & {\rm J}1427-4206   \\
HD~143006       & 2022-04-20 & 47 & 30 & 15-500 & 0.5 & 5.2 & {\rm J}1553-2422 & {\rm J}1517-2422  \\
                & 2022-04-21 & 41 & 30 & 15-500 & 0.5 & 4.7 & {\rm J}1551-1755 & {\rm J}1427-4206   \\
                & 2023-05-01 & 46 & 47 & 15-2517 & 0.09 & 1.8 & {\rm J}1517-2422 & {\rm J}1924-2914   \\
                & 2023-05-07 & 46 & 47 & 15-2517 & 0.09 & 1.6 & {\rm J}1517-2422 & {\rm J}1924-2914   \\
                & 2023-05-09 & 43 & 47 & 15-2517 & 0.09 & 1.6 & {\rm J}1517-2422 & {\rm J}1924-2914   \\
                & 2023-05-10 & 44 & 47 & 15-2517 & 0.09 & 1.6 & {\rm J}1517-2422 & {\rm J}1924-2914   \\
                & 2023-05-11 & 44 & 47 & 15-2517 & 0.09 & 1.5 & {\rm J}1517-2422 & {\rm J}1427-4206   \\
                & 2023-05-11 & 44 & 47 & 15-2517 & 0.09 & 1.5 & {\rm J}1517-2422 & {\rm J}1924-2914   \\
                & 2023-05-12 & 43 & 47 & 15-2517 & 0.09 & 1.4 & {\rm J}1517-2422 & {\rm J}1256-0547   \\
RX{\rm J}1604.3-2130\,A   & 2021-10-25 & 8 & 47 & 9--35 & 3.68 & 19.3 & {\rm J}1553-2422 & {\rm J}1517-2422  \\
                    & 2021-10-25 & 8 & 47 & 9--35 & 3.68 & 19.3 & {\rm J}1625-2527 & {\rm J}1924-2914  \\
                    & 2021-10-30 & 9 & 47 & 9--45 & 3.34 & 19.1 & {\rm J}1625-2527 & {\rm J}1256-0547  \\
                    & 2021-11-09 & 9 & 47 & 9--45 & 3.34 & 19.3 & {\rm J}1625-2527 & {\rm J}1256-0547  \\
                    & 2022-01-29 & 42 & 30 & 15--500 & 0.43 & 5.0 & {\rm J}1551-1755 & {\rm J}1427-4206  \\
                    & 2022-04-20 & 48 & 30 & 15--500 & 0.45 & 5.1 & {\rm J}1553-2422 & {\rm J}1517-2422  \\
                    & 2022-07-12 & 41 & 47 & 15--2517 & 0.11 & 1.7 & {\rm J}1553-2422 & {\rm J}1517-2422  \\
                    & 2022-07-12 & 44 & 47 & 15--2617 & 0.10 & 1.7 & {\rm J}1553-2422 & {\rm J}1517-2422  \\
                    & 2022-07-13 & 47 & 47 & 15--2617 & 0.09 & 1.7 & {\rm J}1553-2422 & {\rm J}1517-2422  \\
RX{\rm J}1615.3-3255  & 2021-11-09 & 10 & 47 & 9--45 & 3.3 & 19.3 & {\rm J}1626-2951 & {\rm J}1427-4206  \\
                & 2022-01-06 & 10 & 47 & 9--45 & 3.3 & 19.3 & {\rm J}1626-2951 & {\rm J}1427-4206  \\
                & 2022-03-26 & 10 & 47 & 9--45 & 3.3 & 19.3 & {\rm J}1626-2951 & {\rm J}1256-0547  \\
                & 2022-04-17 & 46 & 31 & 15--500 & 0.45 & 5.0 & {\rm J}1610-3958 & {\rm J}1517-2422  \\
                & 2022-04-18 & 47 & 31 & 15--500 & 0.45 & 5.0 & {\rm J}1610-3958 & {\rm J}1517-2422  \\
                & 2022-07-26 & 43 & 47 & 15--2617 & 0.09 & 1.7 & {\rm J}1624-3213 & {\rm J}1427-4206  \\
                & 2022-07-26 & 43 & 47 & 15--2617 & 0.09 & 1.7 & {\rm J}1553-2422 & {\rm J}1517-2422   \\
                & 2022-08-05 & 46 & 47 & 15--1301 & 0.17 & 2.3 & {\rm J}1553-2422 & {\rm J}1517-2422   \\
                & 2022-08-07 & 41 & 47 & 15--1301 & 0.16 & 2.1 & {\rm J}1553-2422 & {\rm J}1517-2422  \\
V4046~Sgr       & 2021-10-24 & 10 & 47 & 9--45 & 3.3 & 19.3 & {\rm J}1826-3650 & 	{\rm J}1924-2914  \\
                & 2021-10-28 & 9  & 47 & 9--45 & 3.3 & 19.1 & {\rm J}1826-3650 & 	{\rm J}1924-2914    \\
                & 2021-10-28 & 9  & 47 & 9--45 & 3.3 & 19.1 & {\rm J}1826-3650 & 	{\rm J}1924-2914    \\
                & 2021-11-07 & 10 & 47 & 9--45 & 3.3 & 19.3 & {\rm J}1826-3650 & 	{\rm J}1924-2914   \\
                & 2022-04-18 & 47 & 31 & 15--500 & 0.45 & 5.0 & {\rm J}1733-3722 & {\rm J}1924-2914  \\
                & 2022-04-19 & 48 & 31 & 15--500 & 0.45 & 4.9 & {\rm J}1733-3722 & {\rm J}1924-2914   \\
                & 2022-07-12 & 41 & 47 & 15--2517 & 0.11 & 1.7 & {\rm J}1733-3722 & {\rm J}1924-2914   \\
                & 2022-07-12 & 41 & 47 & 15--2517 & 0.11 & 1.7 & {\rm J}1733-3722 & {\rm J}1924-2914     \\
                & 2022-07-13 & 41 & 47 & 15--2617 & 0.09 & 1.7 & {\rm J}1733-3722 & {\rm J}1924-2914   \\
                & 2022-07-13 & 41 & 47 & 15--2617 & 0.09 & 1.7 & {\rm J}1733-3722 & {\rm J}1924-2914    \\
RX{\rm J}1842.9-3532  & 2022-04-19 & 49 & 31 & 15--500 & 0.45 & 5.0 & {\rm J}1826-3650 & {\rm J}1924-2914 \\
                & 2022-04-20 & 46 & 31 & 15--500 & 0.46 & 5.3 & {\rm J}1826-3650 & {\rm J}1924-2914  \\
                & 2022-07-19 & 44 & 48 & 15--2617 & 0.09 & 1.5 & {\rm J}1826-3650 & {\rm J}1924-2914  \\
                & 2022-07-19 & 44 & 48 & 15--2617 & 0.09 & 1.5 & {\rm J}1826-3650 & {\rm J}1924-2914  \\
                & 2022-07-24 & 41 & 48 & 41--2617 & 0.09 & 1.3 & {\rm J}1826-3650 & {\rm J}1924-2914  \\
                & 2022-07-25 & 41 & 48 & 15--2617 & 0.09 & 1.6 & {\rm J}1826-3650 & {\rm J}1924-2914  \\
                & 2022-07-25 & 41 & 48 & 15--2617 & 0.09 & 1.6 & {\rm J}1826-3650 & {\rm J}1924-2914  \\
RX{\rm J}1852.3-3700  & 2022-04-18 & 47 & 32 & 15--500 & 
                0.45 & 5.0 & {\rm J}1937-3958& {\rm J}1924-2914 \\
                & 2022-04-19 & 48 & 32 & 15--500 &
                0.45 & 4.9 & {\rm J}1937-3958& {\rm J}1924-2914  \\
                & 2022-07-13 & 45 & 48 & 15--2617 &
                0.09 & 1.7 & {\rm J}1937-3958& {\rm J}1924-2914  \\
                & 2022-07-17 & 42 & 48 & 15--2617 &
                0.11 & 1.7 & {\rm J}1937-3958& {\rm J}1924-2914  \\
                & 2022-07-18 & 44 & 48 & 15--2617 &
                0.09 & 1.5 & {\rm J}1937-3958& {\rm J}1924-2914  \\
                & 2022-07-18 & 44 & 48 & 15--2617 &
                0.09 & 1.5 & {\rm J}1937-3958& {\rm J}1924-2914 \\
\enddata
\end{deluxetable*}
\vspace{-2em}

\section{Details of exoALMA Images}
\label{sec:appendix_images}
Table~\ref{tab:images_cont} contains the relevant information of the continuum images provided with the exoALMA data release.

\startlongtable
\begin{deluxetable*}{cccccc}
\tablecaption{Details of all continuum images included in the exoALMA data release. \label{tab:images_cont}}
\tabletypesize{\footnotesize}
\tablehead{
\colhead{Source} & \colhead{Beam Size} & \colhead{JvM $\epsilon$} & \colhead{Theoretical RMS} & \colhead{RMS Achieved} & \colhead{Peak SNR} \\
\colhead{} & \colhead{\footnotesize ($\arcsec{}$)} & \colhead{} & \colhead{\footnotesize (mJy bm$^{-1}$)} & \colhead{\footnotesize ($\mu$Jy bm$^{-1}$)} & \colhead{} 
}
\decimals
\startdata
DM~Tau      &  0.092 $\rm{x}$ 0.077  &  0.356    & 17.1  & 26.6  & 128.6 \\
AA~Tau      &  0.095 $\rm{x}$ 0.075  &  0.498    & 21.9  & 23.8  & 196.7 \\
LkCa~15     &  0.103 $\rm{x}$ 0.078  &  0.502    & 15.3  & 20.3  & 230.9 \\
HD~34282    &  0.087 $\rm{x}$ 0.070  &  0.538    & 17.0  & 22.6  & 270.1 \\
MWC~758     &  0.155 $\rm{x}$ 0.106  &  0.490    & 28.7  & 31.6  & 350.5 \\
CQ~Tau      &  0.111 $\rm{x}$ 0.077  &  0.490    & 24.4  & 25.5  & 539.9 \\
SY~Cha      &  0.130 $\rm{x}$ 0.097  &  0.538    & 30.5  & 30.7  & 90.6 \\
PDS~66      &  0.140 $\rm{x}$ 0.110  &  0.504    & 25.4  & 26.0  & 1232.5 \\
HD~135344B  &  0.163 $\rm{x}$ 0.130  &  0.570    & 21.2  & 23.4  & 705.5 \\
HD~143006   &  0.143 $\rm{x}$ 0.103  &  0.525    & 24.7  & 25.3  & 265.3 \\
RX{\rm J}1604.3-2130      &  0.143 $\rm{x}$ 0.117  &  0.463    & 16.1  & 23.0  & 217.4 \\
RX{\rm J}1615.3-3255      &  0.182 $\rm{x}$ 0.157  &  0.551    & 11.7  & 19.1  & 936.5 \\
V4046~Sgr   &  0.137 $\rm{x}$ 0.119  &  0.503    & 14.8  & 19.7  & 596.0 \\
RX{\rm J}1842.9-3532      &  0.144 $\rm{x}$ 0.114  &  0.487    & 22.5  & 23.4  & 296.6 \\
RX{\rm J}1852.3-3700      &  0.148 $\rm{x}$ 0.113  &  0.565    & 18.9  & 19.7  & 413.5 \\
\enddata
\end{deluxetable*}
\vspace{-2em}

Table~\ref{tab:images_cube_fiducial} contains the relevant information of the `fiducial' line cube images provided with the exoALMA data release.

\startlongtable
\begin{deluxetable*}{cccccccc}
\tablecaption{Details of all `fiducial' line cube images included in the exoALMA data release. \label{tab:images_cube_fiducial}}
\tabletypesize{\footnotesize}
\tablehead{
\colhead{Source} & \colhead{Molecule} & \colhead{Transition} & \colhead{Channel Spacing} & \colhead{Beam Size} & \colhead{JvM $\epsilon$} & \colhead{RMS Achieved} & \colhead{Peak SNR} \\
\colhead{} & \colhead{} & \colhead{(J' - J'')} & \colhead{(m s$^{-1}$)} & \colhead{\footnotesize ($\arcsec{}$)} & \colhead{} & \colhead{\footnotesize (mJy bm$^{-1}$)} & \colhead{} 
}
\decimals
\startdata
DM Tau & CO & 3~-~2 & 100 & 0.15~$\times$~0.15 & 0.154 & 3.14 & 32.05 \\
    & $^{13}$CO & 3~-~2 & 100 & 0.15~$\times$~0.15 & 0.211 & 3.61 & 15.65 \\
    & CS & 7~-~6 & 200 & 0.15~$\times$~0.15 & 0.149 & 1.74 & 10.64 \\
AA Tau & CO & 3~-~2 & 100 & 0.15~$\times$~0.15 & 0.219 & 3.49 & 29.98 \\
    & $^{13}$CO & 3~-~2 & 100 & 0.15~$\times$~0.15 & 0.233 & 3.68 & 13.34 \\
    & CS & 7~-~6 & 200 & 0.15~$\times$~0.15 & 0.218 & 2.09 & 14.99 \\
LkCa 15 & CO & 3~-~2 & 100 & 0.15~$\times$~0.15 & 0.486 & 2.80 & 37.13 \\
    & $^{13}$CO & 3~-~2 & 100 & 0.15~$\times$~0.15 & 0.493 & 2.79 & 21.38 \\
    & CS & 7~-~6 & 200 & 0.15~$\times$~0.15 & 0.482 & 1.67 & 20.98 \\
HD 34282 & CO & 3~-~2 & 100 & 0.15~$\times$~0.15 & 0.294 & 2.74 & 56.00 \\
    & $^{13}$CO & 3~-~2 & 100 & 0.15~$\times$~0.15 & 0.285 & 3.15 & 31.50 \\
    & CS & 7~-~6 & 200 & 0.15~$\times$~0.15 & 0.293 & 1.67 & 9.55 \\
MWC 758 & CO & 3~-~2 & 100 & 0.15~$\times$~0.15 & 1.0 & 4.39 & 37.73 \\
    & $^{13}$CO & 3~-~2 & 100 & 0.15~$\times$~0.15 & 0.961 & 5.18 & 19.37 \\
    & CS & 7~-~6 & 200 & 0.15~$\times$~0.15 & 0.948 & 2.63 & 6.65 \\
CQ Tau & CO & 3~-~2 & 100 & 0.15~$\times$~0.15 & 0.336 & 2.98 & 49.32 \\
    & $^{13}$CO & 3~-~2 & 100 & 0.15~$\times$~0.15 & 0.436 & 3.25 & 21.76 \\
    & CS & 7~-~6 & 200 & 0.15~$\times$~0.15 & 0.400 & 1.81 & 7.26 \\
SY Cha & CO & 3~-~2 & 100 & 0.15~$\times$~0.15 & 0.780 & 3.91 & 27.13 \\
    & $^{13}$CO & 3~-~2 & 100 & 0.15~$\times$~0.15 & 0.881 & 4.43 & 12.09 \\
    & CS & 7~-~6 & 200 & 0.15~$\times$~0.15 & 0.763 & 2.34 & 9.77 \\
PDS 66 & CO & 3~-~2 & 100 & 0.15~$\times$~0.15 & 0.801 & 3.42 & 31.21 \\
    & $^{13}$CO & 3~-~2 & 100 & 0.15~$\times$~0.15 & 0.910 & 3.67 & 14.53 \\
    & CS & 7~-~6 & 200 & 0.15~$\times$~0.15 & 0.897 & 2.06 & 8.75 \\
HD 135344B & CO & 3~-~2 & 100 & 0.15~$\times$~0.15 & 0.996 & 2.98 & 67.11 \\
    & $^{13}$CO & 3~-~2 & 100 & 0.15~$\times$~0.15 & 1.0 & 3.62 & 32.64 \\
    & CS & 7~-~6 & 200 & 0.15~$\times$~0.15 & 1.0 & 1.87 & 13.22 \\
HD 143006 & CO & 3~-~2 & 100 & 0.15~$\times$~0.15 & 0.714 & 3.12 & 36.96 \\
    & $^{13}$CO & 3~-~2 & 100 & 0.15~$\times$~0.15 & 0.811 & 3.66 & 13.48 \\
    & CS & 7~-~6 & 200 & 0.15~$\times$~0.15 & 0.728 & 1.78 & 11.58 \\
RXJ1604.3-2130 & CO & 3~-~2 & 100 & 0.15~$\times$~0.15 & 0.740 & 3.30 & 49.82 \\
    & $^{13}$CO & 3~-~2 & 100 & 0.15~$\times$~0.15 & 0.914 & 3.93 & 31.15 \\
    & CS & 7~-~6 & 200 & 0.15~$\times$~0.15 & 0.827 & 2.05 & 25.03 \\
RXJ1615.3-3255 & CO & 3~-~2 & 100 & 0.15~$\times$~0.15 & 0.923 & 2.85 & 31.84 \\
    & $^{13}$CO & 3~-~2 & 100 & 0.15~$\times$~0.15 & 1.0 & 3.13 & 19.41 \\
    & CS & 7~-~6 & 200 & 0.15~$\times$~0.15 & 0.884 & 1.72 & 11.07 \\
V4046 Sgr & CO & 3~-~2 & 100 & 0.15~$\times$~0.15 & 0.799 & 2.73 & 60.97 \\
    & $^{13}$CO & 3~-~2 & 100 & 0.15~$\times$~0.15 & 0.770 & 2.71 & 35.21 \\
    & CS & 7~-~6 & 200 & 0.15~$\times$~0.15 & 0.794 & 1.61 & 18.49 \\
RXJ1842.9-3532 & CO & 3~-~2 & 100 & 0.15~$\times$~0.15 & 0.900 & 3.15 & 36.67 \\
    & $^{13}$CO & 3~-~2 & 100 & 0.15~$\times$~0.15 & 0.879 & 3.38 & 13.19 \\
    & CS & 7~-~6 & 200 & 0.15~$\times$~0.15 & 0.916 & 1.90 & 9.14 \\
RXJ1852.3-3700 & CO & 3~-~2 & 100 & 0.15~$\times$~0.15 & 0.964 & 2.90 & 47.13 \\
    & $^{13}$CO & 3~-~2 & 100 & 0.15~$\times$~0.15 & 0.979 & 3.09 & 22.62 \\
    & CS & 7~-~6 & 200 & 0.15~$\times$~0.15 & 0.974 & 1.73 & 26.01 \\
\enddata
\end{deluxetable*}
\vspace{-2em}

Table~\ref{tab:images_cube_highres} contains the relevant information of the `high resolution' line cube images provided with the exoALMA data release.

\startlongtable
\begin{deluxetable*}{cccccccc}
\tablecaption{Details of all `high resolution' line cube images included in the exoALMA data release. Some disks do not have $^{13}$CO high resolution images due to SNR limitations. \label{tab:images_cube_highres}}
\tabletypesize{\footnotesize}
\tablehead{
\colhead{Source} & \colhead{Molecule} & \colhead{Transition} & \colhead{Channel Spacing} & \colhead{Beam Size} & \colhead{JvM $\epsilon$} & \colhead{RMS Achieved} & \colhead{Peak SNR} \\
\colhead{} & \colhead{} & \colhead{(J' - J'')} & \colhead{(m s$^{-1}$)} & \colhead{\footnotesize ($\arcsec{}$)} & \colhead{} & \colhead{\footnotesize (mJy bm$^{-1}$)} & \colhead{} 
}
\decimals
\startdata
DM Tau & CO & 3~-~2 & 200 & 0.091~$\times$~0.084 & 0.489 & 2.36 & 25.70 \\
    & $^{13}$CO & 3~-~2 & 200 & 0.111~$\times$~0.101 & 0.257 & 2.44 & 14.90 \\
AA Tau & CO & 3~-~2 & 100 & 0.085~$\times$~0.083 & 0.651 & 3.58 & 17.89 \\
    & $^{13}$CO & 3~-~2 & 200 & 0.120~$\times$~0.103 & 0.277 & 2.63 & 13.19 \\
LkCa 15 & CO & 3~-~2 & 200 & 0.086~$\times$~0.067 & 0.975 & 2.53 & 14.48 \\
    & $^{13}$CO & 3~-~2 & 200 & 0.097~$\times$~0.079 & 0.725 & 2.40 & 13.10 \\
HD 34282 & CO & 3~-~2 & 200 & 0.073~$\times$~0.060 & 1.000 & 2.58 & 17.45 \\
    & $^{13}$CO & 3~-~2 & 200 & 0.095~$\times$~0.089 & 0.500 & 2.35 & 22.74 \\
MWC 758 & CO & 3~-~2 & 200 & 0.105~$\times$~0.078 & 1.000 & 4.51 & 17.58 \\
    & $^{13}$CO & 3~-~2 & 200 & 0.124~$\times$~0.089 & 0.706 & 3.95 & 14.75 \\
CQ Tau & CO & 3~-~2 & 200 & 0.110~$\times$~0.091 & 0.583 & 2.43 & 44.16 \\
    & $^{13}$CO & 3~-~2 & 200 & 0.111~$\times$~0.091 & 0.630 & 2.66 & 19.87 \\
SY Cha & CO & 3~-~2 & 200 & 0.105~$\times$~0.077 & 0.861 & 3.52 & 15.03 \\
PDS 66 & CO & 3~-~2 & 200 & 0.109~$\times$~0.084 & 0.758 & 2.98 & 16.48 \\
HD 135344B & CO & 3~-~2 & 200 & 0.108~$\times$~0.091 & 0.746 & 2.64 & 38.31 \\
HD 143006 & CO & 3~-~2 & 200 & 0.097~$\times$~0.072 & 1.000 & 3.84 & 14.88 \\
RXJ1604.3-2130 & CO & 3~-~2 & 100 & 0.109~$\times$~0.084 & 0.595 & 3.97 & 20.44 \\
RXJ1615.3-3255 & CO & 3~-~2 & 200 & 0.120~$\times$~0.105 & 0.746 & 2.41 & 24.40 \\
    & $^{13}$CO & 3~-~2 & 200 & 0.130~$\times$~0.114 & 0.734 & 2.52 & 17.80 \\
V4046 Sgr & CO & 3~-~2 & 200 & 0.092~$\times$~0.077 & 0.803 & 3.10 & 22.25 \\
    & $^{13}$CO & 3~-~2 & 200 & 0.104~$\times$~0.101 & 0.777 & 2.50 & 28.14 \\
RXJ1842.9-3532 & CO & 3~-~2 & 200 & 0.110~$\times$~0.084 & 0.685 & 2.71 & 23.79 \\
RXJ1852.3-3700 & CO & 3~-~2 & 200 & 0.106~$\times$~0.075 & 0.935 & 3.16 & 19.54 \\
    & $^{13}$CO & 3~-~2 & 200 & 0.121~$\times$~0.088 & 0.704 & 2.52 & 16.16 \\
\enddata
\end{deluxetable*}
\vspace{-2em}

Table~\ref{tab:images_cube_lowres} contains the relevant information of the `low resolution' line cube images provided with the exoALMA data release.

\startlongtable
\begin{deluxetable*}{cccccccc}
\tablecaption{Details of all `low resolution' line cube images included in the exoALMA data release. \label{tab:images_cube_lowres}}
\tabletypesize{\footnotesize}
\tablehead{
\colhead{Source} & \colhead{Molecule} & \colhead{Transition} & \colhead{Channel Spacing} & \colhead{Beam Size} & \colhead{JvM $\epsilon$} & \colhead{RMS Achieved} & \colhead{Peak SNR} \\
\colhead{} & \colhead{} & \colhead{(J' - J'')} & \colhead{(m s$^{-1}$)} & \colhead{\footnotesize ($\arcsec{}$)} & \colhead{} & \colhead{\footnotesize (mJy bm$^{-1}$)} & \colhead{} 
}
\decimals
\startdata
DM Tau & CO & 3~-~2 & 100 & 0.30~$\times$~0.30 & 0.368 & 3.25 & 88.58 \\
    & $^{13}$CO & 3~-~2 & 100 & 0.30~$\times$~0.30 & 0.320 & 3.35 & 42.26 \\
    & CS & 7~-~6 & 200 & 0.30~$\times$~0.30 & 0.379 & 1.97 & 20.67 \\
AA Tau & CO & 3~-~2 & 100 & 0.30~$\times$~0.30 & 0.463 & 3.74 & 71.13 \\
    & $^{13}$CO & 3~-~2 & 100 & 0.30~$\times$~0.30 & 0.433 & 4.20 & 28.85 \\
    & CS & 7~-~6 & 200 & 0.30~$\times$~0.30 & 0.466 & 2.26 & 29.44 \\
LkCa 15 & CO & 3~-~2 & 100 & 0.30~$\times$~0.30 & 0.582 & 3.06 & 92.91 \\
    & $^{13}$CO & 3~-~2 & 100 & 0.30~$\times$~0.30 & 0.587 & 3.10 & 49.52 \\
    & CS & 7~-~6 & 200 & 0.30~$\times$~0.30 & 0.562 & 1.81 & 43.80 \\
HD 34282 & CO & 3~-~2 & 100 & 0.30~$\times$~0.30 & 0.650 & 3.66 & 114.87 \\
    & $^{13}$CO & 3~-~2 & 100 & 0.30~$\times$~0.30 & 0.602 & 4.02 & 52.93 \\
    & CS & 7~-~6 & 200 & 0.30~$\times$~0.30 & 0.625 & 2.20 & 14.75 \\
MWC 758 & CO & 3~-~2 & 100 & 0.30~$\times$~0.30 & 0.597 & 3.94 & 124.67 \\
    & $^{13}$CO & 3~-~2 & 100 & 0.30~$\times$~0.30 & 0.589 & 4.06 & 68.29 \\
    & CS & 7~-~6 & 200 & 0.30~$\times$~0.30 & 0.606 & 2.42 & 12.75 \\
CQ Tau & CO & 3~-~2 & 100 & 0.30~$\times$~0.30 & 0.437 & 3.33 & 98.18 \\
    & $^{13}$CO & 3~-~2 & 100 & 0.30~$\times$~0.30 & 0.393 & 3.43 & 41.04 \\
    & CS & 7~-~6 & 200 & 0.30~$\times$~0.30 & 0.441 & 1.99 & 9.21 \\
SY Cha & CO & 3~-~2 & 100 & 0.30~$\times$~0.30 & 0.746 & 5.88 & 50.03 \\
    & $^{13}$CO & 3~-~2 & 100 & 0.30~$\times$~0.30 & 0.483 & 4.12 & 28.42 \\
    & CS & 7~-~6 & 200 & 0.30~$\times$~0.30 & 0.706 & 3.35 & 15.80 \\
PDS 66 & CO & 3~-~2 & 100 & 0.30~$\times$~0.30 & 0.521 & 3.17 & 88.29 \\
    & $^{13}$CO & 3~-~2 & 100 & 0.30~$\times$~0.30 & 0.495 & 3.05 & 37.42 \\
    & CS & 7~-~6 & 200 & 0.30~$\times$~0.30 & 0.468 & 1.80 & 19.89 \\
HD 135344B & CO & 3~-~2 & 100 & 0.30~$\times$~0.30 & 0.516 & 2.56 & 196.46 \\
    & $^{13}$CO & 3~-~2 & 100 & 0.30~$\times$~0.30 & 0.722 & 2.84 & 111.24 \\
    & CS & 7~-~6 & 200 & 0.30~$\times$~0.30 & 0.536 & 1.52 & 35.99 \\
HD 143006 & CO & 3~-~2 & 100 & 0.30~$\times$~0.30 & 0.664 & 3.32 & 104.79 \\
    & $^{13}$CO & 3~-~2 & 100 & 0.30~$\times$~0.30 & 0.588 & 3.37 & 38.92 \\
    & CS & 7~-~6 & 200 & 0.30~$\times$~0.30 & 0.677 & 1.90 & 28.11 \\
RXJ1604.3-2130 & CO & 3~-~2 & 100 & 0.30~$\times$~0.30 & 0.572 & 2.92 & 176.20 \\
    & $^{13}$CO & 3~-~2 & 100 & 0.30~$\times$~0.30 & 0.560 & 3.08 & 120.73 \\
    & CS & 7~-~6 & 200 & 0.30~$\times$~0.30 & 0.525 & 1.69 & 101.10 \\
RXJ1615.3-3255 & CO & 3~-~2 & 100 & 0.30~$\times$~0.30 & 0.498 & 2.25 & 107.22 \\
    & $^{13}$CO & 3~-~2 & 100 & 0.30~$\times$~0.30 & 0.574 & 2.29 & 68.39 \\
    & CS & 7~-~6 & 200 & 0.30~$\times$~0.30 & 0.536 & 1.32 & 30.72 \\
V4046 Sgr & CO & 3~-~2 & 100 & 0.30~$\times$~0.30 & 0.503 & 2.45 & 154.60 \\
    & $^{13}$CO & 3~-~2 & 100 & 0.30~$\times$~0.30 & 0.491 & 2.31 & 81.50 \\
    & CS & 7~-~6 & 200 & 0.30~$\times$~0.30 & 0.529 & 1.41 & 58.26 \\
RXJ1842.9-3532 & CO & 3~-~2 & 100 & 0.30~$\times$~0.30 & 0.471 & 2.68 & 112.54 \\
    & $^{13}$CO & 3~-~2 & 100 & 0.30~$\times$~0.30 & 0.445 & 2.80 & 41.99 \\
    & CS & 7~-~6 & 200 & 0.30~$\times$~0.30 & 0.564 & 1.74 & 19.43 \\
RXJ1852.3-3700 & CO & 3~-~2 & 100 & 0.30~$\times$~0.30 & 0.542 & 2.49 & 149.51 \\
    & $^{13}$CO & 3~-~2 & 100 & 0.30~$\times$~0.30 & 0.544 & 2.43 & 79.42 \\
    & CS & 7~-~6 & 200 & 0.30~$\times$~0.30 & 0.653 & 1.58 & 72.46 \\
\enddata
\end{deluxetable*}
\vspace{-2em}

\vspace{5mm}
\facilities{ALMA}


\software{\texttt{CASA} v6.2.1.7 \citep{CASA2022}, \texttt{analysisUtils} \citep{Hunter2023}, \texttt{emcee} \citep{Foreman-Mackey2013,Foreman-Mackey2019}, \texttt{numpy} \citep{numpy}, \texttt{scipy} \citep{Virtanen2020}, \texttt{matplotlib} \citep{Hunter2007}, \texttt{astropy} \citep{astropy2013,astropy2018,astropy2022}, \texttt{JupyterNotebook} \citep{jupyternootbok}, \texttt{numba} \citep{numba}, \texttt{reduction\_utils} scripts by the DSHARP and MAPS ALMA Large Programs \citep{Andrews_ea_2018,Czekala_ea_2021}.}




\bibliography{cal_image}{}
\bibliographystyle{aasjournal}



\end{document}